\documentclass[showpacs,nofootinbib,showkeys,eqsecnum,prd,aps,preprint]{revtex4-1}
\usepackage{amsfonts,amsmath,amssymb,bm,graphicx}

\begin{document}

\title{{\large Scattering and pair creation by a constant electric field
between two capacitor plates}}
\author{S.~P.~Gavrilov${}^{a,c}$}
\email{gavrilovsergeyp@yahoo.com}
\author{D.~M.~Gitman${}^{a,b,d}$}
\email{gitman@if.usp.br}
\date{\today }

\begin{abstract}
Using the quantum field theory approach developed in Phys. Rev. D. \textbf{93%
}, 045002 (2016), we consider particle scattering and vacuum instability in
the so-called $L$-constant electric field, which is a constant electric
field confined between two capacitor plates separated by a finite distance $%
L $. We obtain and analyze special sets of stationary solutions of the Dirac
and Klein-Gordon equations with the $L$-constant electric field. Then, we
represent probabilities of particle scattering and characteristics of the
vacuum instability (related to pair creation) in terms of the introduced
solutions. From exact formulas, we derive asymptotic expressions for the
differential mean numbers, for the total mean number of created particles,
and for the vacuum-to-vacuum transition probability. Using the equivalence
principle, we demonstrate that the distributions of particles created by the
$L$-constant electric field and the gravitational field of a black hole have
a similar thermal structure.
\end{abstract}


\affiliation{${}^{a}$Department of Physics, Tomsk State University, 634050 Tomsk, Russia;\\
${}^{b}$P.N. Lebedev Physical Institute, 53 Leninskiy prospect, 119991 Moscow,
Russia;\\
${}^{c}$Department of General and Experimental Physics, Herzen State
Pedagogical University of Russia, Moyka embankment 48, 191186
St.~Petersburg, Russia\\
${}^{d}$Institute of Physics, University of S\~{a}o Paulo, CP 66318, CEP
05315-970 S\~{a}o Paulo, SP, Brazil}

\maketitle

\section{Introduction\label{S1}}

The effect of particle creation by strong electromagnetic and gravitational
fields has a pure quantum nature. Depending on the structure of such
external backgrounds, different approaches have been proposed for
calculating the effect. Initially, the effect of particle creation was
considered for time-dependent external electric fields that are switched on
and off at the initial and final time instants, respectively.\ We call such
external fields the $t$-electric potential steps. Scattering, particle
creation from the vacuum, and particle annihilation by the $t$-electric
potential steps has been considered in the framework of the relativistic
quantum mechanics, see Refs.~\cite{Nikis70a,Nikis79,GMR85}; a more complete
list of relevant publications can be found in Refs.~\cite{ruffini,GelTan15}.
At present it is well understood that only an adequate quantum field theory
(QFT) with the corresponding external background may consistently describe
this effect and possible accompanying{\large \ }processes. In the framework
of such a theory, particle creation is related to a violation of the vacuum
stability with time. Backgrounds (external fields) that may violate the
vacuum stability have to be able to produce nonzero work when interacting
with the corresponding particles. In quantum electrodynamics (QED), these
are electriclike electromagnetic fields. A general formulation of QED with $%
t $-electric potential steps{\large \ }was developed in Refs. \cite{Gitman}.
However, there exist many physically interesting situations where external
backgrounds are not formally switched off at the time infinity, the
corresponding backgrounds formally being not $t$-electric potential steps.
As an example, we may point out time-independent nonuniform electric fields
that are concentrated in restricted space areas. The latter fields represent
a kind of spatial (or, as we call them, conditionally) $x$-electric
potential steps for charged particles. The $x$-electric potential steps can
also create particles from the vacuum; the Klein paradox is closely related
to this process \cite{Klein27,Sauter31a,Sauter-pot}.

Approaches for treating quantum effects in the explicitly time-dependent
external fields are not directly applicable to the $x$-electric potential
steps. Some heuristic calculations of particle creation by $x$-electric
potential steps in the framework of the relativistic quantum mechanics with
qualitative discussion from the point of view of QFT were first presented by
Nikishov in Refs.~\cite{Nikis79,Nikis70b}. In our recent article \cite%
{GavGit15}, we presented a consistent formulation of QED with $x$-electric
potential steps quantizing the Dirac and the Klein-Gordon (scalar) fields in
the presence of such steps in terms of adequate \textrm{in-} and \textrm{out}%
-particles. We developed a nonperturbative calculation technique for
different quantum processes such as scattering, reflection, and
electron-positron pair creation. As in the case of QED with $t$-electric
potential steps this technique essentially uses special sets of exact
solutions of the Dirac equation with the corresponding external field of $x$%
-electric potential steps. The cases when such solutions\textrm{\ }can be
found explicitly (analytically){\large \ }are{\large \ }called exactly
solvable cases. In QED with $t$-electric potential steps there exist a few
exactly solvable cases. The first (conditionally) is the so-called $T$%
-constant electric field, which is a uniform electric field that efficiently
acts during a sufficiently large but finite time $T$. Quantum processes
including particle creation in the latter field were studied in detail in
Refs. ~\cite{BagGitSh75,GavG96a,GG06-08,GavG08,GavGitY12} and used for a
number of applications \cite{GavGT06,GavGDS14}. One can also point out other
exactly solvable cases, such as the Sauter-like electric field $\,$\cite%
{NarNik70} (see also Refs.~\cite{GavG96a,DunHal98}), and an exponentially
decreasing electric field \cite{ExpoF}. In the recently constructed QED with
$x$-electric potential steps an important exactly solvable case of the
Sauter field $E(x)=E\cosh ^{-2}\left( x/L_{\mathrm{S}}\right) $ was
considered in detail in Ref.~\cite{GavGit15} as an illustration of the
general theory. In the present article we consider the second exactly
solvable case of $x$-electric potential steps, namely the so-called $L$%
-constant electric field. Such a step represents a constant electric field
situated between two planes $x=x_{\mathrm{L}}=\mathrm{const}$ and $x=x_{%
\mathrm{R}}=\mathrm{const}$, where $x_{\mathrm{R}}-x_{\mathrm{L}}=L$, is
uniform there and directed along the axis $x$, see Sec. \ref{S2}. In fact,
this is a uniform electric field confined between two capacitor plates
separated by a finite distance $L$. Some heuristic calculations of the
particle creation effect by such a field in the framework of the
relativistic quantum mechanics were presented in Ref.~\cite{WongW88}{\large .%
} The $L$-constant electric field is an analog of the $T$-constant field. We
stress that the $T$-constant and $L$-constant electric fields describe
different physical situations in the general case. However, in the limiting
case, when both $T\rightarrow \infty $ and $L\rightarrow \infty $, they
represent the same uniform constant electric field, which is obviously an
idealization. Nevertheless, such a field allows exact solutions and has been
frequently used in various calculations in QED, particularly in the
pioneering work of Schwinger \cite{Schwi51} (see \cite{Dunn04} for a
review). These calculations are relatively simple due to the{\large \ }%
translational symmetry of the field; however they always contain divergences
related to the infinite duration of the field action and to the infinite
volume of the consideration. In this respect calculations in the $T$%
-constant and $L$-constant electric fields can be considered as a kind of
different regularization in the case of the constant uniform electric field.

The study of the vacuum instability{\large \ }in the presence of potential
steps -- in particular, particle creation in the $L$-constant electric field
-- is quite important for various applications. For example, it is important
in the study of particle emission from black holes and quark and neutron
stars, due to a close relation between particle creation by strong
electrostatic potentials and the Unruh effect; see, e.g., \cite%
{ruffini,CHM08} for reviews. The corresponding limiting case of a constant
uniform electric field has many similarities with the case of the de Sitter
background, see, e.g., Refs.~\cite{AndMot14,AkhmP15} and references therein.
Recent progress in laser physics allows one to hope that particle creation
effect will be experimentally observed in the near future in laboratory
conditions (see Refs.~\cite{Dun09} for a review). Methods of QFT with strong
potential steps are currently being developed in condensed matter physics,
and particle creation by external fields has become an observable effect in
physics of graphene and similar nanostructures (say, in topological
insulators and\ Weyl semimetals); this area is currently under intense
development, see the reviews \cite{dassarma,VafVish14}. In particular, the
particle creation effect is crucial for understanding the conductivity of
graphene, especially in the so-called nonlinear regime. Electron-hole pair
creation (which is an analog of the electron-positron pair creation from the
vacuum) was recently observed in graphene by its indirect influence on the
graphene conductivity \cite{Van+etal10}. Possible experimental
configurations for testing the pair creation by a linear step of finite
length were proposed in \cite{allor}. The inhomogeneity of the field becomes
important in achieving extreme field strengths.

In this article, using the general approach developed in Ref. \cite{GavGit15}%
, we consider the particle scattering and the vacuum instability on a
quasilinear $x$-electric potential step corresponding to an $L$-constant
electric field. We use notation from and final formulas of the latter work.
In Sec. \ref{S2}, we obtain and analyze special sets of stationary solutions
of the Dirac and Klein-Gordon equations with the $L$-constant electric
field. In Sec. \ref{S3}, we represent probabilities of particle scattering
and characteristics of the vacuum instability (related to the pair creation)
in terms of the introduced solutions. From exact formulas, we derive
asymptotic expressions for differential mean numbers, the total mean number
of created particles, and the vacuum-to-vacuum transition probability in the
case of a small-gradient field. In the first part of the discussion, Sec. %
\ref{SS4.1}, we consider length and time scales of small-gradient fields and
show that quantum effects in an $L$-constant field are quite representative
for a large class of small-gradient electric fields. In the second part of
the discussion, Sec. \ref{SS4.2}, we show that the distribution of particles
created by a strong electrostatic inhomogeneous fields of a small gradient
-- in particular, by the $L$-constant field -- can be written in a general
Hawking-like thermal form, in which the Hawking temperature is reproduced
exactly. Section \ref{S5} contains our Conclusion. In Appendix \ref{Ap}, we
describe briefly basic elements of QED with $x$-electric potential steps. In
Appendix \ref{Ap2}, we list some useful properties of the Weber parabolic
cylinder functions (WPCFs). In what follows we use the system of units where
$\hslash =c=1$.

\section{In- and out-solutions in an $L$-constant electric field\label{S2}}

\subsection{Dirac equation}

\begin{figure}[tbp]
\centering\includegraphics[scale=.6]{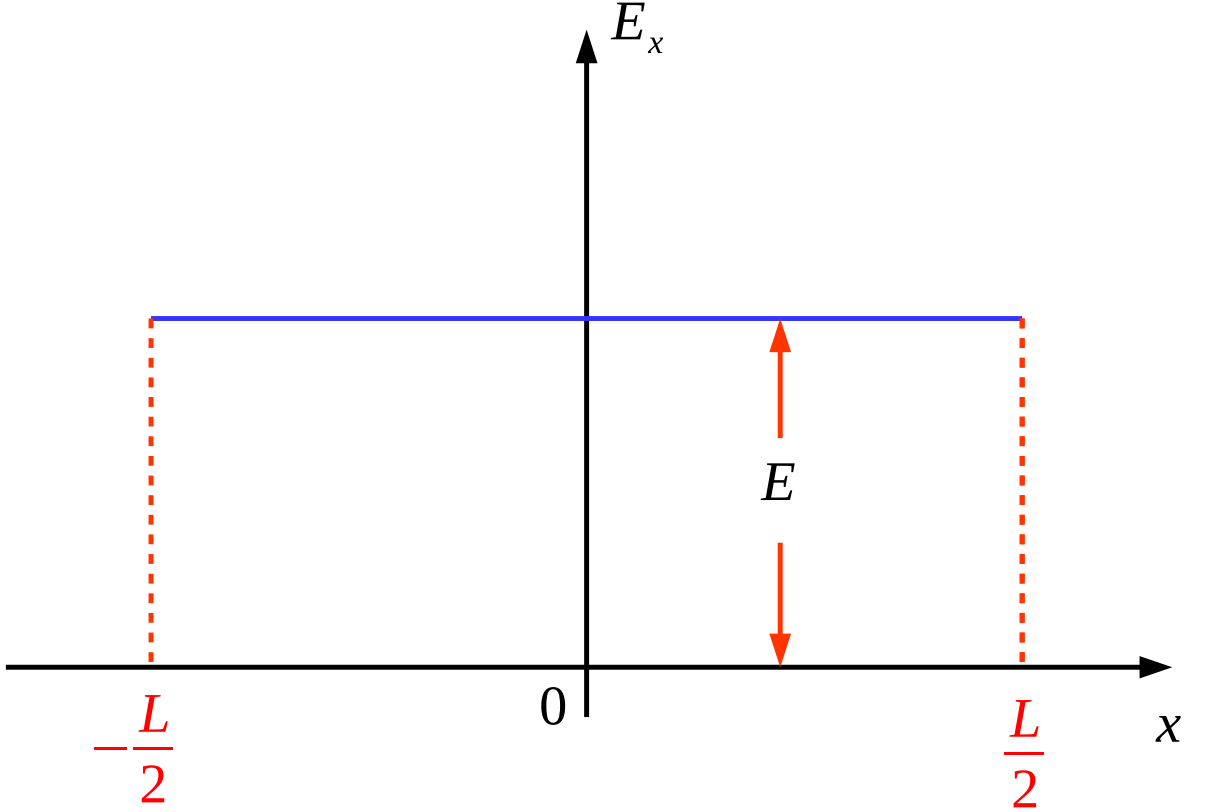}
\caption{$L$-constant electric field}
\label{1}
\end{figure}

Let us consider QED with an $x$-electric potential step, which is an $L$%
-constant electric field. The latter electromagnetic field consists of a
pure electric field $\mathbf{E}$ (the corresponding magnetic field $\mathbf{B%
}$ is zero) of the form\footnote{%
We recall that our system is placed in the $d=D+1$ dimensional Minkowski
spacetime parametrized by the coordinates $X=\left( X^{\mu },\ \mu
=0,1,\ldots ,D\right) =\left( t,\mathbf{r}\right) $, $X^{0}=t$, $\ \ \mathbf{%
r}=\left( X^{1},\ldots ,X^{D}\right) $. It consists of a Dirac field $\psi
\left( X\right) $ interacting with an external electromagnetic field $A^{\mu
}(X)$ in the form of a $x$-electric potential step.} $\mathbf{E}\left(
X\right) =\mathbf{E}\left( x\right) =\left( E_{x}\left( x\right)
,0,...,0\right) .\ $The electric field $E\left( x\right) $ has the form%
\begin{equation*}
E\left( x\right) =E=\mathrm{const}>0,\;x\in S_{\mathrm{int}}=\left( x_{%
\mathrm{L}},x_{\mathrm{R}}\right) ;\quad E\left( x\right) =0\,,\;x\in S_{%
\mathrm{L}}=\left( -\infty ,x_{\mathrm{L}}\right] ,\ x\in S_{\mathrm{R}}=%
\left[ x_{\mathrm{R}},\infty \right) ,
\end{equation*}%
and we choose that $x_{\mathrm{L}}=-L/2$ and $x_{\mathrm{R}}=L/2$. The plot
of the field is shown in Fig.~\ref{1}.

We assume that the basic Dirac particle is an electron with the mass $m$ and
the charge $-e$, $e>0$, and the positron is its antiparticle. The electric
field under consideration accelerates the electrons along the $x$ axis in
the negative direction and the positrons along the $x$ axis in the positive
direction.

Potentials of the corresponding electromagnetic field $A^{\mu }\left(
X\right) $ can be chosen as%
\begin{equation}
A^{\mu }\left( X\right) =\left( A^{0}\left( x\right) ,A^{j}=0,\
j=1,2,...,D\right) ,\ \ x=X^{1},  \label{2.3}
\end{equation}%
[so that $E\left( x\right) =-A_{0}^{\prime }\left( x\right) $]with a
linearly growing potential $A^{0}\left( x\right) $ on an interval $x\in S_{%
\mathrm{int}}$ of the length $L=x_{\mathrm{R}}-x_{\mathrm{L}}$, which is
constant out of this interval. The potential energy of an electron in the
electric field under consideration is $U\left( x\right) =-eA_{0}\left(
x\right) $,%
\begin{equation}
U\left( x\right) =\left\{
\begin{array}{ll}
U_{\mathrm{L}}=eEx_{\mathrm{L}}, & x\in S_{\mathrm{L}} \\
eEx, & x\in S_{\mathrm{int}} \\
U_{\mathrm{R}}=eEx_{\mathrm{R}}, & x\in S_{\mathrm{R}}%
\end{array}%
\right. .  \label{L1}
\end{equation}%
The magnitude of the $x$-electric step under consideration is
\begin{equation}
\mathbb{U=}U_{\mathrm{R}}-U_{\mathrm{L}}=eEL>0.  \label{L2}
\end{equation}

In $d$ dimensions, the Dirac field $\psi \left( X\right) $\ is a column with
$2^{\left[ d/2\right] }$ components ( in what follows we call it just a
spinor), and $\gamma ^{\mu }$ are $2^{\left[ d/2\right] }\times 2^{\left[ d/2%
\right] }$ gamma matrices; see, e.g., Ref.~\cite{BraWe35},
\begin{equation*}
\lbrack \gamma ^{\mu },\gamma ^{\nu }]_{+}=2\eta ^{\mu \nu },\;\eta ^{\mu
\nu }=\mathrm{diag}(\underbrace{1,-1,\ldots ,-1}_{d}),\;\mu ,\nu =0,1,\ldots
,D.
\end{equation*}

The classical Dirac field $\psi \left( X\right) $ satisfies the Dirac
equation with the $L$-constant electric field,
\begin{equation}
i\partial _{0}\psi \left( X\right) =\hat{H}\psi \left( X\right) ,\ \ \hat{H}%
=\gamma ^{0}\left( -i\gamma ^{j}\partial _{j}+m\right) +U\left( x\right) ,
\label{e1}
\end{equation}%
where $\hat{H}$ is the one-particle Dirac Hamiltonian .

Let us consider stationary solutions of the Dirac equation (\ref{e1}) having
the following form%
\begin{eqnarray}
&&\psi _{n}\left( X\right) =\exp \left( -ip_{0}t+i\mathbf{p}_{\bot }\mathbf{r%
}_{\bot }\right) \psi _{n}\left( x\right) ,\ \ X=\left( t,x,\mathbf{r}_{\bot
}\right) ,\;\ n=(p_{0},\mathbf{p}_{\bot },\sigma ),  \notag \\
&&\psi _{n}\left( x\right) =\left\{ \gamma ^{0}\left[ p_{0}-U\left( x\right) %
\right] -\gamma ^{1}\hat{p}_{x}-\boldsymbol{\gamma }_{\bot }\mathbf{p}_{\bot
}+m\right\} \phi _{n}(x),  \notag \\
&&\mathbf{r}_{\bot }=\left( X^{2},\ldots ,X^{D}\right) ,\ \mathbf{p}_{\bot
}=\left( p^{2},\ldots ,p^{D}\right) ,\ \boldsymbol{\gamma }_{\bot }=\left(
\gamma ^{2},...,\gamma ^{D}\right) ,\ \hat{p}_{x}=-i\partial _{x},
\label{e2}
\end{eqnarray}%
where $\psi _{n}\left( x\right) $ and $\phi _{n}(x)$ are spinors that depend
on $x$ alone. In fact these are stationary states with the energy $p_{0}$
and with definite momenta $\mathbf{p}_{\bot }$ in the directions
perpendicular to the axis $x$. Substituting (\ref{e2}) into Dirac equation (%
\ref{e1}) (i.e., partially squaring the Dirac equation), we obtain a
second-order differential equation for the spinor $\phi _{n}(x)$,
\begin{equation}
\left\{ \hat{p}_{x}^{2}-i\gamma ^{0}\gamma ^{1}U^{\prime }\left( x\right) -%
\left[ p_{0}-U\left( x\right) \right] ^{2}+\mathbf{p}_{\bot
}^{2}+m^{2}\right\} \phi _{n}(x)=0.  \label{e2.1}
\end{equation}%
We separate spinning variables by the substitution
\begin{equation}
\phi _{n}(x)=\phi _{n}^{\left( \chi \right) }(x)=\varphi _{n}\left( x\right)
v_{\chi ,\sigma },\   \label{2.33}
\end{equation}%
where $v_{\chi ,\sigma }$ with $\chi =\pm 1$ and $\sigma =(\sigma
_{1},\sigma _{2},\dots ,\sigma _{\lbrack d/2]-1})$, $\sigma _{s}=\pm 1$, is
a set of constant orthonormalized spinors, satisfying the following equations%
\begin{equation}
\gamma ^{0}\gamma ^{1}v_{\chi ,\sigma }=\chi v_{\chi ,\sigma },\;\;v_{\chi
,\sigma }^{\dag }v_{\chi ^{\prime },\sigma ^{\prime }}=\delta _{\chi ,\chi
^{\prime }}\delta _{\sigma ,\sigma ^{\prime }}.  \label{e2a}
\end{equation}%
The quantum numbers $\chi $ and $\sigma _{s}$ describe a spin polarization
and provide a convenient parametrization of the solutions. Since in ($1+1$)
and $\left( 2+1\right) $ dimensions ($d=2,3$) there are no any spin degrees
of freedom, the quantum numbers $\sigma $ are absent. Then, scalar functions
$\varphi _{n}\left( x\right) $ have to obey the second-order differential
equation%
\begin{equation}
\left\{ \hat{p}_{x}^{2}-i\chi U^{\prime }\left( x\right) -\left[
p_{0}-U\left( x\right) \right] ^{2}+\mathbf{p}_{\bot }^{2}+m^{2}\right\}
\varphi _{n}\left( x\right) =0.  \label{e3}
\end{equation}

In $d$ dimensions, for any given set $p_{0},\mathbf{p}_{\bot },$ there
exists only $J_{(d)}=2^{[d/2]-1}$ different spin states. The projection
operator, which is situated inside the brackets $\left\{ \cdots \right\} $
in Eq.~(\ref{e2}), does not commute with the matrix $\gamma ^{0}\gamma ^{1}$
and, consequently, transforms $\phi _{n}^{\left( \chi \right) }(x)$ with a
given $\chi $ to a linear superposition of functions $\phi _{n}^{\left(
+1\right) }(x)$ and $\phi _{n^{\prime }}^{\left( -1\right) }(x)$ with
indices $n$ and $n^{\prime }$ corresponding to the same $p_{0},\mathbf{p}%
_{\bot }$. In $d\geq 4$\ dimensions{\large \ }this projection operator also
does not commute with the matrix $\Xi =i\boldsymbol{\gamma }_{\bot }\mathbf{p%
}_{\bot }/\left\vert \mathbf{p}_{\bot }\right\vert $. Assuming that $\sigma
_{1}=\pm 1$ is an eigenvalue of $\Xi $, one can see that solutions (\ref{e2}%
), which differ only by values of $\chi $ and $\sigma _{1}$, are linearly
dependent. Let us denote by $\psi _{n_{\pm }}^{\left( \chi \right) }\left(
x\right) $ solutions of the Dirac equation with quantum numbers $n_{\pm
}=(p_{0},\mathbf{p}_{\bot },\sigma _{\pm })$, $\sigma _{\pm }=(\pm 1,\sigma
_{2},\dots ,\sigma _{\lbrack d/2]-1})$. Then, for example, using existing
relations between solutions (\ref{e2}) for $x\in S_{\mathrm{L}}$, one can
easily verify that%
\begin{equation}
\left[ \pi _{0}\left( \mathrm{L}\right) -p^{\mathrm{L}}\right] \psi _{n_{\pm
}}^{\left( +1\right) }\left( x\right) =\left( \pm i\left\vert \mathbf{p}%
_{\bot }\right\vert +m\right) \psi _{n_{\mp }}^{\left( -1\right) }\left(
x\right) .  \label{ld}
\end{equation}%
The case of $\left( 1+1\right) $ dimensions follows from Eq.~(\ref{ld}) at $%
\left\vert \mathbf{p}_{\bot }\right\vert =0$ and $n_{\pm }=n_{\mp }=p_{0}$,
assuming $m\neq 0$. Note that in the case of $\left( 2+1\right) $
dimensions, there are two nonequivalent representations for the $\gamma $
matrices,
\begin{equation*}
\gamma ^{0}=\sigma ^{3},\;\gamma ^{1}=i\sigma ^{2},\;\gamma ^{2}=-i\sigma
^{1}s,\;\;s=\pm 1\,,
\end{equation*}%
which correspond to different fermion species. In this case, following the
same logic, one can see that%
\begin{equation*}
\left[ \pi _{0}\left( \mathrm{L}\right) -p^{\mathrm{L}}\right] \psi
_{n}^{\left( +1\right) }\left( x\right) =\left( isp^{2}+m\right) \psi
_{n}^{\left( -1\right) }\left( x\right) ,\;\ n=(p_{0},p^{2}).
\end{equation*}%
That is why it is sufficient to work only with solutions corresponding to
one of the values of $\chi $. In what follows, we fix the quantum number $%
\chi $ in a certain way.

\subsection{In- and out-solutions}

In what follows, we use solutions of the Dirac equation denoted as $%
_{\;\zeta }\psi _{n}\left( X\right) $ and $^{\;\zeta }\psi _{n}\left(
X\right) ,\ \zeta =\pm \ ,$ with special left and right asymptotics at $x\in
S_{\mathrm{L}}$ and $x\in S_{\mathrm{R}}.$ Such solutions have the form (\ref%
{e2}) with the functions $\varphi _{n}\left( x\right) $ denoted as $%
_{\;\zeta }\varphi _{n}\left( x\right) $ or $^{\;\zeta }\varphi _{n}\left(
x\right) $, respectively. The latter functions satisfy Eq.~(\ref{e3}) and
the following asymptotic conditions:
\begin{eqnarray}
&&_{\;\zeta }\varphi _{n}\left( x\right) =\ _{\zeta }\mathcal{N}\exp \left[
ip^{\mathrm{L}}\left( x-x_{\mathrm{L}}\right) \right] ,\ x\in S_{\mathrm{L}},
\notag \\
&&^{\;\zeta }\varphi _{n}\left( x\right) =\ ^{\zeta }\mathcal{N}\exp \left[
ip^{\mathrm{R}}\left( x-x_{\mathrm{R}}\right) \right] ,\ x\in S_{\mathrm{R}},
\notag \\
&&p^{\mathrm{L}}=\zeta \sqrt{\left[ \pi _{0}\left( \mathrm{L}\right) \right]
^{2}-\pi _{\bot }^{2}},\ \ p^{\mathrm{R}}=\zeta \sqrt{\left[ \pi _{0}\left(
\mathrm{R}\right) \right] ^{2}-\pi _{\bot }^{2}},\ \zeta =\pm \ ,  \notag \\
&&\pi _{0}\left( \mathrm{L/R}\right) =p_{0}-U_{\mathrm{L/R}},\ \ \pi _{\bot
}=\sqrt{\mathbf{p}_{\bot }^{2}+m^{2}}.  \label{L3}
\end{eqnarray}%
Thus, the solutions $_{\;\zeta }\psi _{n}\left( X\right) $ and $^{\;\zeta
}\psi _{n}\left( X\right) $ asymptotically describe particles with given
real momenta $p^{\mathrm{L/R}}$ along the $x$ axis. The factors$\ _{\zeta }%
\mathcal{N}$ and $\ ^{\zeta }\mathcal{N}$ are normalization constants with
respect to conditions (\ref{c3}) given in Appendix \ref{Ap},
\begin{eqnarray}
\ _{\zeta }\mathcal{N} &\mathcal{=}&_{\zeta }CY,\;\;\ ^{\zeta }\mathcal{N}=\
^{\zeta }CY,\;\;Y=\left( V_{\bot }T\right) ^{-1/2},  \notag \\
\ _{\zeta }C &=&\left[ 2\left\vert p^{\mathrm{L}}\right\vert \left\vert \pi
_{0}\left( \mathrm{L}\right) -\chi p^{\mathrm{L}}\right\vert \right]
^{-1/2},\;\ ^{\zeta }C=\left[ 2\left\vert p^{\mathrm{R}}\right\vert
\left\vert \pi _{0}\left( \mathrm{R}\right) -\chi p^{\mathrm{R}}\right\vert %
\right] ^{-1/2}.  \label{e8b}
\end{eqnarray}

Since $p_{0}$ is the total energy of a particle, we interpret $\pi
_{0}\left( \mathrm{R}\right) $ and $\pi _{0}\left( \mathrm{L}\right) $ as
sum of its asymptotic kinetic and the rest energies in the regions $S_{%
\mathrm{R}}$ and $S_{\mathrm{L}}$, respectively. We call the quantity $\pi
_{\bot }$ the transversal energy.

Nontrivial solutions $\ ^{\zeta }\psi _{n}\left( X\right) $ exist only for
quantum numbers $n$ that obey the relation%
\begin{equation}
\left[ \pi _{0}\left( \mathrm{R}\right) \right] ^{2}>\pi _{\bot
}^{2}\Longleftrightarrow \left\{
\begin{array}{l}
\pi _{0}\left( \mathrm{R}\right) >\pi _{\bot } \\
\pi _{0}\left( \mathrm{R}\right) <-\pi _{\bot }%
\end{array}%
\right. ,  \label{2.62d}
\end{equation}%
whereas nontrivial solutions $_{\zeta }\psi _{n}\left( X\right) $ exist only
for quantum numbers $n$ that obey the relation%
\begin{equation}
\left[ \pi _{0}\left( \mathrm{L}\right) \right] ^{2}>\pi _{\bot
}^{2}\Longleftrightarrow \left\{
\begin{array}{l}
\pi _{0}\left( \mathrm{L}\right) >\pi _{\bot } \\
\pi _{0}\left( \mathrm{L}\right) <-\pi _{\bot }%
\end{array}%
\right. .  \label{2.6c}
\end{equation}

We distinguish two types of electric steps, noncritical and critical, by
their magnitudes as follows:%
\begin{equation}
\mathbb{U}=\left\{
\begin{array}{l}
\mathbb{U}<\mathbb{U}_{c}=2m,\ \mathrm{noncritical\ step} \\
\mathbb{U}>\mathbb{U}_{c},\ \mathrm{critical\ step}%
\end{array}%
\right. .  \label{2.63}
\end{equation}

In the case of noncritical steps, the vacuum is stable, see Ref. \cite%
{GavGit15}. We are interested in the critical steps where there is
electron-positron pair production from the vacuum.

In the case of critical steps of any form, and in particular the step under
consideration, there exist five ranges of quantum numbers $n$, \emph{\ }$%
\Omega _{k}$\emph{, }$k=1,...,5$, where the introduced solutions have
similar forms, and physical processes with particles have similar
interpretation; see Ref. \cite{GavGit15}.

(i) In the range \emph{\ }$\Omega _{1}$, where $p_{0}\geq U_{\mathrm{R}}+\pi
_{\bot }$, and the range $\emph{\ }\Omega _{5}$, where $p_{0}\leq U_{\mathrm{%
L}}-\pi _{\bot }$, the number of particles is conserved. In the range $%
\Omega _{1}$ there exist only incoming [$\ _{+}\psi _{n}\left( X\right) $ or
$\ ^{-}\psi _{n}\left( X\right) $] and outgoing [ $_{-}\psi _{n}\left(
X\right) $ or $\ ^{+}\psi _{n}\left( X\right) $] electrons, whereas in the
range $\emph{\ }\Omega _{5},$\ there exist only incoming [$\ _{-}\psi
_{n}\left( X\right) $ and $\ ^{+}\psi _{n}\left( X\right) $] and outgoing [$%
\ _{+}\psi _{n}\left( X\right) $ and $\ ^{-}\psi _{n}\left( X\right) $]
positrons. In these ranges there exist only scattering and the reflection of
the particles. Particle creation is impossible in these ranges.

(ii) In the range $\Omega _{2}$, where%
\begin{eqnarray*}
U_{\mathrm{R}}-\pi _{\bot } &<&p_{0}<U_{\mathrm{R}}+\pi _{\bot },\ \ \pi
_{0}\left( \mathrm{L}\right) >\pi _{\bot }\ \mathrm{if\;}2\pi _{\bot }\leq
\mathbb{U}, \\
U_{\mathrm{L}}+\pi _{\bot } &<&p_{0}<U_{\mathrm{R}}+\pi _{\bot }\ \mathrm{%
if\;}2\pi _{\bot }>\mathbb{U},
\end{eqnarray*}%
any solution has zero right asymptotic, which means that we deal with
standing waves of the form%
\begin{equation}
\psi _{n}\left( X\right) =_{\ +}\psi _{n}\left( X\right) e^{+i\theta _{n}}+\
_{-}\psi _{n}\left( X\right) e^{-i\theta _{n}}\ .  \label{2.46b}
\end{equation}%
Here, similar to the range\emph{\ }$\Omega _{1}$, there exist only electrons
that are subjected to the total\emph{\ }reflection. In the range $\Omega
_{4} $, where%
\begin{eqnarray*}
U_{\mathrm{L}}-\pi _{\bot } &<&p_{0}<U_{\mathrm{L}}+\pi _{\bot },\ \ \pi
_{0}\left( \mathrm{R}\right) <-\pi _{\bot }\ \mathrm{if\;}2\pi _{\bot }\leq
U, \\
U_{\mathrm{L}}-\pi _{\bot } &<&p_{0}<U_{\mathrm{R}}-\pi _{\bot }\ \mathrm{%
if\;}2\pi _{\bot }>U,
\end{eqnarray*}%
any solution has zero left asymptotic, which means that we deal with
standing waves of the form%
\begin{equation}
\psi _{n}\left( X\right) =\ ^{+}\psi _{n}\left( X\right) e^{+i\theta _{n}}+\
^{-}\psi _{n}\left( X\right) e^{-i\theta _{n}}\ .  \label{2.46a}
\end{equation}%
Here, similar to $\Omega _{5}$, there exist only positrons that are also
subjected to the total\emph{\ }reflection. The number of particles in $%
\Omega _{2}$ and $\Omega _{4}$ is conserved.

(iii) The range \emph{\ }$\Omega _{3}$, where $U_{\mathrm{L}}+\pi _{\bot
}\leq p_{0}\leq U_{\mathrm{R}}-\pi _{\bot }$, exists only for transversal
momenta that satisfy the inequality{\large \ }$2\pi _{\bot }\leq \mathbb{U}$%
. Here the QFT description of quantum processes is essential. Its brief
description is given in Appendix \ref{Ap}. In this range, there exist
\textrm{in}- and \textrm{out}-electrons that can be situated only to the
left of the step, and \textrm{in}- and \textrm{out}-positrons that can be
situated only to the right of the step. In the range $\Omega _{3}$, all the
partial vacua are unstable, and particle creation is possible. These pairs
consist of \textrm{out}-electrons and \textrm{out}-positrons that appear on
the left and on the right of the step and move there{\large \ }to the left
and to the right, respectively. At the same time, the \textrm{in}-electrons
that move\emph{\ }to the step from the left are subjected to the total
reflection. After\emph{\ }being reflected, they move to the left of the step
as \textrm{out}-electrons. Similarly, the \textrm{in}-positrons that move%
\emph{\ } to the step from the right are subjected to the total reflection.\
After being reflected, they move to the right of the step as \textrm{out}%
-positrons.

It is assumed that each pair of solutions $_{\zeta }\psi _{n}\left( X\right)
$ and $^{\zeta }\psi _{n}\left( X\right) $ with given quantum numbers $n\in
\Omega _{1}\cup \Omega _{3}\cup \Omega _{5}$ is complete in the space of
solutions with the corresponding $n.$ Because Eq.~(\ref{c3}), given in
Appendix \ref{Ap}, the mutual decompositions of such solutions have the form%
\begin{eqnarray}
\eta _{\mathrm{L}}\ ^{\zeta }\psi _{n}\left( X\right) &=&\ _{+}\psi
_{n}\left( X\right) g\left( _{+}\left\vert ^{\zeta }\right. \right) -\
_{-}\psi _{n}\left( X\right) g\left( _{-}\left\vert ^{\zeta }\right. \right)
,  \notag \\
\eta _{\mathrm{R}}\ _{\zeta }\psi _{n}\left( X\right) &=&\ ^{+}\psi
_{n}\left( X\right) g\left( ^{+}\left\vert _{\zeta }\right. \right) -\
^{-}\psi _{n}\left( X\right) g\left( ^{-}\left\vert _{\zeta }\right. \right)
,  \label{rel1}
\end{eqnarray}%
where the decomposition coefficients $g$ are
\begin{equation}
\left( \ _{\zeta }\psi _{n},\ ^{\zeta ^{\prime }}\psi _{n^{\prime }}\right)
_{x}=\delta _{n,n^{\prime }}g\left( \ _{\zeta }\left\vert ^{\zeta ^{\prime
}}\right. \right) ,\ \ g\left( \ ^{\zeta ^{\prime }}\left\vert _{\zeta
}\right. \right) =g\left( \ _{\zeta }\left\vert ^{\zeta ^{\prime }}\right.
\right) ^{\ast },\ \ n\in \Omega _{1}\cup \Omega _{3}\cup \Omega _{5}\ .
\label{c12}
\end{equation}%
These coefficients satisfy the following unitary relations%
\begin{eqnarray}
&&\left\vert g\left( _{-}\left\vert ^{+}\right. \right) \right\vert
^{2}=\left\vert g\left( _{+}\left\vert ^{-}\right. \right) \right\vert
^{2},\;\left\vert g\left( _{+}\left\vert ^{+}\right. \right) \right\vert
^{2}=\left\vert g\left( _{-}\left\vert ^{-}\right. \right) \right\vert
^{2},\;\frac{g\left( _{+}\left\vert ^{-}\right. \right) }{g\left(
_{-}\left\vert ^{-}\right. \right) }=\frac{g\left( ^{+}\left\vert
_{-}\right. \right) }{g\left( ^{+}\left\vert _{+}\right. \right) },  \notag
\\
&&\left\vert g\left( _{+}\left\vert ^{-}\right. \right) \right\vert
^{2}-\left\vert g\left( _{+}\left\vert ^{+}\right. \right) \right\vert
^{2}=-\eta _{\mathrm{L}}\eta _{\mathrm{R}}.  \label{UR}
\end{eqnarray}

For $x\in S_{\mathrm{int}}$, Eqs.~(\ref{e3}) can be written in the form%
\begin{equation}
\left[ \frac{\mathrm{d}^{2}}{\mathrm{d}\xi ^{2}}+\xi ^{2}+\mathrm{i}\chi
-\lambda \right] \varphi _{n}\left( x\right) =0,  \label{L4}
\end{equation}%
where
\begin{equation}
\xi =\frac{eEx-p_{0}}{\sqrt{eE}},\ \ \lambda =\frac{\pi _{\bot }^{2}}{eE}.
\label{L5}
\end{equation}%
The general solution of Eq.~(\ref{L4}) is completely determined by an
appropriate pair of the linearly independent Weber parabolic cylinder
functions (WPCFs): either $D_{\rho }[(1-\mathrm{i})\xi ]$ and $D_{-1-\rho
}[(1+\mathrm{i})\xi ],$ or $D_{\rho }[-(1-\mathrm{i})\xi ]$ and $D_{-1-\rho
}[-(1+\mathrm{i})\xi ]$, where $\rho =-\mathrm{i}\lambda /2-\left( 1+\chi
\right) /2$. Then, taking into account Eq.~(\ref{rel1}), the functions $\
_{-}\varphi _{n}\left( x\right) $ and $\ ^{+}\varphi _{n}\left( x\right) $
can be presented in the form%
\begin{eqnarray}
\ _{-}\varphi _{n}\left( x\right) &=&Y%
\begin{cases}
\ _{-}C\exp \left[ -i\left\vert p^{\mathrm{L}}\right\vert \left( x-x_{%
\mathrm{L}}\right) \right] , & x\in S_{\mathrm{L}} \\
\ _{-}C\left\{ a_{1}D_{\rho }[-(1-i)\xi ]+a_{2}D_{-1-\rho }[-(1+i)\xi
]\right\} , & x\in S_{\mathrm{int}} \\
\eta _{\mathrm{R}}\left\{ g\left( ^{+}\left\vert _{-}\right. \right) \
^{+}C\exp \left[ i\left\vert p^{\mathrm{R}}\right\vert \left( x-x_{\mathrm{R}%
}\right) \right] -g\left( ^{-}\left\vert _{-}\right. \right) \ ^{-}C\exp %
\left[ -i\left\vert p^{\mathrm{R}}\right\vert \left( x-x_{\mathrm{R}}\right) %
\right] \right\} , & x\in S_{\mathrm{R}}%
\end{cases}%
;  \notag \\
\ ^{+}\varphi _{n}\left( x\right) &=&Y%
\begin{cases}
\eta _{\mathrm{L}}\left\{ g(_{+}|^{+})\ _{+}C\exp \left[ i\left\vert p^{%
\mathrm{L}}\right\vert \left( x-x_{\mathrm{L}}\right) \right] -g(_{-}|^{+})\
_{-}C\exp \left[ -i\left\vert p^{\mathrm{L}}\right\vert \left( x-x_{\mathrm{L%
}}\right) \right] \right\} , & x\in S_{\mathrm{L}} \\
\ ^{+}C\left\{ a_{1}^{\prime }D_{\rho }[(1-i)\xi ]+a_{2}^{\prime }D_{-1-\rho
}[(1+i)\xi ]\right\} , & x\in S_{\mathrm{int}} \\
\ ^{+}C\exp \left[ i\left\vert p^{\mathrm{R}}\right\vert \left( x-x_{\mathrm{%
R}}\right) \right] , & x\in S_{\mathrm{R}}%
\end{cases}
\label{L6}
\end{eqnarray}%
on the whole axis $x.$ The functions $\ _{-}\varphi _{n}\left( x\right) $
and $\ ^{+}\varphi _{n}\left( x\right) $ and their derivatives satisfy the
following gluing conditions:
\begin{equation}
\ _{-}^{+}\varphi _{n}(x_{\mathrm{L/R}}-0)=\ _{-}^{+}\varphi _{n}(x_{\mathrm{%
L/R}}+0),\quad \partial _{x}\ _{-}^{+}\varphi _{n}(x_{\mathrm{L/R}%
}-0)=\partial _{x}\ _{-}^{+}\varphi _{n}(x_{\mathrm{L/R}}+0).  \label{L7}
\end{equation}%
Note that the following relations hold: $\left\vert p^{\mathrm{L}%
}\right\vert /\sqrt{eE}=\sqrt{\xi _{1}^{2}-\lambda }$ and $\left\vert p^{%
\mathrm{R}}\right\vert /\sqrt{eE}=\sqrt{\xi _{2}^{2}-\lambda }$, where $\xi
_{1}=\left. \xi \right\vert _{x=x_{\mathrm{L}}}$, \ $\xi _{2}=\left. \xi
\right\vert _{x=x_{\mathrm{R}}}$.

A formal transition to the Klein-Gordon case can be done by setting $\chi =0$
and $\eta _{\mathrm{L}}=\eta _{\mathrm{R}}=1$\ in Eqs. (\ref{L6}). In this
case $n=(p_{0},\mathbf{p}_{\bot })$, and normalization factors are%
\begin{equation*}
\ _{\zeta }\mathcal{N}\mathcal{=}_{\zeta }CY,\;\;\ ^{\zeta }\mathcal{N}=\
^{\zeta }CY,\;\;\ _{\zeta }C=\left\vert 2p^{\mathrm{L}}\right\vert
^{-1/2},\;\ ^{\zeta }C=\left\vert 2p^{\mathrm{R}}\right\vert ^{-1/2}.
\end{equation*}

Using Eq.~(\ref{L7}) and the Wronskian determinant of WPCFs, we find the
coefficients $a_{j}$ and $a_{j}^{\prime },$
\begin{eqnarray}
a_{j} &=&-\frac{(-1)^{j}}{\sqrt{2}}\exp \left[ \frac{i\pi }{2}\left( \rho +%
\frac{1}{2}\right) \right] \sqrt{\xi _{1}^{2}-\lambda }f_{j}^{(-)}(\xi _{1}),
\notag \\
a_{j}^{\prime } &=&-\frac{(-1)^{j}}{\sqrt{2}}\exp \left[ \frac{i\pi }{2}%
\left( \rho +\frac{1}{2}\right) \right] \sqrt{\xi _{2}^{2}-\lambda }%
f_{j}^{(+)}(\xi _{2}),\quad j=1,2,  \label{L8}
\end{eqnarray}%
where%
\begin{eqnarray}
f_{1}^{(\pm )}(\xi ) &=&\left( 1\pm \frac{i}{\sqrt{\xi ^{2}-\lambda }}\frac{d%
}{d\xi }\right) D_{-\rho -1}\left[ \pm (1+i)\xi \right] ,  \notag \\
f_{2}^{(\pm )}(\xi ) &=&\left( 1\pm \frac{i}{\sqrt{\xi ^{2}-\lambda }}\frac{d%
}{d\xi }\right) D_{\rho }\left[ \pm (1-i)\xi \right] .  \label{L9}
\end{eqnarray}%
They can be used to determine the coefficients $g\left( {}_{\pm
}|{}^{+}\right) $ and $g\left( {}^{\pm }|{}_{-}\right) $. It should be noted
that we need to know explicitly only the coefficients $g\left(
{}_{-}|{}^{+}\right) $ and $g\left( {}^{+}|{}_{-}\right) $, which are%
\begin{eqnarray}
&&g\left( {}^{+}|{}_{-}\right) =\eta _{\mathrm{R}}AB\exp \left[ \left( \rho
+1/2\right) i\pi /2\right] ,\ \ g\left( {}_{-}|{}^{+}\right) =\eta _{\mathrm{%
L}}A^{\prime }B^{\prime }\exp \left[ \left( \rho +1/2\right) i\pi /2\right] ,
\notag \\
&&A=\sqrt{\frac{\sqrt{\xi _{2}^{2}-\lambda }\sqrt{\xi _{1}^{2}-\lambda }%
\left\vert \xi _{2}+\chi \sqrt{\xi _{2}^{2}-\lambda }\right\vert }{%
8\left\vert \xi _{1}-\chi \sqrt{\xi _{1}^{2}-\lambda }\right\vert }},\ \
B=f_{1}^{(-)}(\xi _{1})f_{2}^{(-)}(\xi _{2})-f_{2}^{(-)}(\xi
_{1})f_{1}^{(-)}(\xi _{2}),  \notag \\
&&A^{\prime }=\sqrt{\frac{\sqrt{\xi _{2}^{2}-\lambda }\sqrt{\xi
_{1}^{2}-\lambda }\left\vert \xi _{1}-\chi \sqrt{\xi _{1}^{2}-\lambda }%
\right\vert }{8\left\vert \xi _{2}+\chi \sqrt{\xi _{2}^{2}-\lambda }%
\right\vert }},\ \ B^{\prime }=f_{1}^{(+)}(\xi _{1})f_{2}^{(+)}(\xi
_{2})-f_{2}^{(+)}(\xi _{1})f_{1}^{(+)}(\xi _{2}).  \label{L10}
\end{eqnarray}%
One can see that coefficients (\ref{L10}) obey the relations
\begin{equation}
\left. g\left( {}^{+}|{}_{-}\right) \right\vert _{p_{0}\rightarrow
-p_{0}}=-\eta _{\mathrm{L}}\eta _{\mathrm{R}}g\left( {}_{-}|{}^{+}\right) .
\label{L11}
\end{equation}%
From these relations, one can conclude that $\left\vert g\left(
{}_{-}|{}^{+}\right) \right\vert $ is an even function of the energy $p_{0}$%
, transversal momenta $\mathbf{p}_{\bot }$ and and does not depend on a spin
polarization.

In the Klein-Gordon case, the coefficients $g$ are
\begin{eqnarray}
&&g\left( {}^{+}|{}_{-}\right) =\exp \left( \lambda \pi /4\right) A_{\mathrm{%
sc}}\left. B\right\vert _{\chi =0},\ \ g\left( {}_{-}|{}^{+}\right) =\exp
\left( \lambda \pi /4\right) A_{\mathrm{sc}}\left. B^{\prime }\right\vert
_{\chi =0},  \notag \\
&&A_{\mathrm{sc}}=\left( \frac{1}{8}\sqrt{\xi _{2}^{2}-\lambda }\sqrt{\xi
_{1}^{2}-\lambda }\right) ^{1/2},\ \   \label{L12}
\end{eqnarray}%
where $B$ and $B^{\prime }$ are given by Eqs. (\ref{L10}). They satisfy the
unitary relations (\ref{UR}) in which we have to set $\eta _{\mathrm{L}%
}=\eta _{\mathrm{R}}=1$.

As follows from Eqs.~(\ref{L10}) and (\ref{L12}), if either $\left\vert p^{%
\mathrm{R}}\right\vert $ or $\left\vert p^{\mathrm{L}}\right\vert $ tends to
zero, one of the following limits holds true :%
\begin{equation}
\left\vert g\left( _{-}\left\vert ^{+}\right. \right) \right\vert ^{-2}\sim
\sqrt{\xi _{2}^{2}-\lambda }\rightarrow 0,\ \ \left\vert g\left(
_{-}\left\vert ^{+}\right. \right) \right\vert ^{-2}\sim \sqrt{\xi
_{1}^{2}-\lambda }\rightarrow 0,\ \ \forall \lambda \neq 0.  \label{L15}
\end{equation}%
These properties are essential for the justification of \textrm{in}- and
\textrm{out}-particle interpretation in the general construction \cite%
{GavGit15}.

The modulus $\left\vert g\left( _{-}\left\vert ^{+}\right. \right)
\right\vert $ in the form ~(\ref{L10}) and (\ref{L12}) was obtained in the
course of heuristic calculations in the framework of the relativistic
quantum mechanics in Ref. \cite{WongW88}.

\section{Scattering and creation of particles \label{S3}}

\subsection{Ranges of stable vacuum}

We know that in the ranges $\Omega _{i}$, $i=1,2,4,5$ the partial vacua are
stable. Let us start with discussion of formulas obtained for these ranges.
In the ranges \emph{\ }$\Omega _{2}$ and $\Omega _{4}$, a particle is
subjected to the total\emph{\ }reflection. In the adjacent ranges,\emph{\ }$%
\Omega _{1}$ and $\Omega _{5}$, a particle can be reflected and transmitted.
For example, in the range $\Omega _{1},$ total $\tilde{R}$ and relative $R$
amplitudes of an electron reflection, and total $\tilde{T}$\ and relative $T$
amplitudes of an electron transmission can be presented as the following
matrix elements%
\begin{eqnarray}
R_{+,n} &=&\tilde{R}_{+,n}c_{v}^{-1},\ \tilde{R}_{+,n}=\langle 0,\mathrm{out}%
|\ _{-}a_{n}(\mathrm{out})\ _{+}a_{n}^{\dag }(\mathrm{in})|0,\mathrm{in}%
\rangle ,  \notag \\
T_{+,n} &=&\tilde{T}_{+,n}c_{v}^{-1},\ \tilde{T}_{+,n}=\langle 0,\mathrm{out}%
|\ ^{+}a_{n}(\mathrm{out})\ _{+}a_{n}^{\dag }(\mathrm{in})|0,\mathrm{in}%
\rangle ,  \notag \\
R_{-,n} &=&\tilde{R}_{-,n}c_{v}^{-1},\ \tilde{R}_{-,n}=\langle 0,\mathrm{out}%
|\ ^{+}a_{n}(\mathrm{out})\ ^{-}a_{n}^{\dag }(\mathrm{in})|0,\mathrm{in}%
\rangle ,  \notag \\
T_{-,n} &=&\tilde{T}_{-,n}c_{v}^{-1},\ \tilde{T}_{-,n}=\langle 0,\mathrm{out}%
|_{-}a_{n}(\mathrm{out})\ ^{-}a_{n}^{\dag }(\mathrm{in})|0,\mathrm{in}%
\rangle ,  \label{L13}
\end{eqnarray}%
where state vectors in the corresponding Fock space and the vacuum-to-vacuum
transition amplitude $c_{v}$ are defined in Appendix \ref{Ap}. The relative
reflection $\left\vert R_{\zeta ,n}\right\vert ^{2}$ and transmission $%
\left\vert T_{\zeta ,n}\right\vert ^{2}$ probabilities satisfy the relation
\textrm{\ }%
\begin{equation}
\left\vert T_{\zeta ,n}\right\vert ^{2}=1-\left\vert R_{\zeta ,n}\right\vert
^{2},\;\;\left\vert R_{\zeta ,n}\right\vert ^{2}=\left[ 1+\left\vert g\left(
_{-}\left\vert ^{+}\right. \right) \right\vert ^{-2}\right] ^{-1},\ \zeta
=\pm \ .  \label{L14}
\end{equation}%
Similar expressions can be derived for positron amplitudes in the range $%
\Omega _{5}$. In particular, relation (\ref{L14}) holds true literally for
the positrons in the range $\Omega _{5}$.

It is clear that $\left\vert R_{\zeta ,n}\right\vert ^{2}\leq 1$. This
result may be interpreted as QFT justification of the rules of
time-independent potential scattering theory in the ranges $\Omega _{1}$ and
$\Omega _{5}$. Amplitudes of Klein-Gordon particle reflection and
transmission in the ranges $\Omega _{i}$, $i=1,2,4,5$ have the same form as
in the Dirac particle case with coefficients $g$ given by the corresponding
inner product. Substituting the coefficients $g$ given by Eqs.~(\ref{L10})
or (\ref{L12}) into relations.~(\ref{L14}), one can find explicitly
reflection and transmission probabilities in the $L$-constant field.

The limits (\ref{L15}) imply the following properties of the coefficients $%
\left\vert g\left( _{-}\left\vert ^{+}\right. \right) \right\vert $:

(i) $\left\vert g\left( _{-}\left\vert ^{+}\right. \right) \right\vert
^{-2}\rightarrow 0$ in the range $\Omega _{1}$ if $n$ tends to the boundary
with the range $\Omega _{2}$ ($\left\vert p^{\mathrm{R}}\right\vert
\rightarrow 0$);

(ii) $\left\vert g\left( _{-}\left\vert ^{+}\right. \right) \right\vert
^{-2}\rightarrow 0$ in the range $\Omega _{5}$ if $n$ tends to the boundary
with the range $\Omega _{4}$ ($\left\vert p^{\mathrm{L}}\right\vert
\rightarrow 0$).

Thus, in the above cases the relative reflection probabilities $\left\vert
R_{\zeta ,n}\right\vert ^{2}$ tend to the unity; i.e., they are continuous
functions of the quantum numbers $n$ on the boundaries. In addition, it
follows from Eqs.~(\ref{L10}) and (\ref{L12}) that $\left\vert g\left(
_{-}\left\vert ^{+}\right. \right) \right\vert ^{2}\rightarrow 0$ and,
therefore, $\left\vert R_{\zeta ,n}\right\vert ^{2}\rightarrow 0$ as $%
p_{0}\rightarrow \pm \infty $, as of course it must be.

\subsection{Klein zone}

The range $\Omega _{3}$ which is called the Klein zone is of special
interest due to the vacuum instability. We recall, as it follows from the
general consideration \cite{GavGit15}, that if in the range $\Omega _{3}$
there exists an \textrm{in}-particle, it will be subjected to the total\emph{%
\ }reflection. For example, it follows from Eq.~(\ref{cq31}), given in
Appendix \ref{Ap}, that the probability of reflection of a particle with
given quantum numbers $n$, under the condition that all other partial vacua
remain vacua, is $P(+|+)_{n,n}P_{v}^{-1}p_{v}^{n}=1$. In the Dirac case, the
presence of an \textrm{in}-particle with a given $n\in \Omega _{3}$
disallows the pair creation from the vacuum in this state due to the Pauli
principle. Of course, pairs of bosons can be created from the vacuum in any
already-occupied states.

The differential mean numbers of created pairs have the form $N_{n}^{\mathrm{%
cr}}=\left\vert g\left( _{-}\left\vert ^{+}\right. \right) \right\vert ^{-2}$%
; see Eq.~(\ref{7.5}), where $g\left( _{-}\left\vert ^{+}\right. \right) $
is given by Eq.~(\ref{L10}) for Dirac particles and by Eq.~(\ref{L12}) for
Klein-Gordon particles. Note that dimensionless parameters $\lambda $ and $%
\xi $ entering these expressions satisfy the condition%
\begin{equation}
\sqrt{\lambda }\leq \xi _{2},\;\;\xi _{1}\leq -\sqrt{\lambda }\ ,
\label{L16b}
\end{equation}%
which are, in fact, consequences of the definition of the range \emph{\ }$%
\Omega _{3}$.

From properties (\ref{L15}), one finds that $N_{n}^{\mathrm{cr}}\rightarrow
0 $ if $n$ tends to the boundary with either the range $\Omega _{2}$ $\left(
\sqrt{\xi _{2}^{2}-\lambda }\rightarrow 0\right) $ or the range $\Omega _{4}$
$\left( \sqrt{\xi _{1}^{2}-\lambda }\rightarrow 0\right) $; in the latter
ranges, the vacuum is stable.

One can see that absolute values of $\sqrt{\xi _{1}^{2}-\lambda }$ and $%
\sqrt{\xi _{2}^{2}-\lambda }$ are related as follows:
\begin{equation}
0\leq \left\vert \sqrt{\xi _{2}^{2}-\lambda }-\sqrt{\xi _{1}^{2}-\lambda }%
\right\vert \leq \omega ,\;\;\omega =\left[ \sqrt{eE}L\left( \sqrt{eE}L-2%
\sqrt{\lambda }\right) \right] ^{1/2}.  \label{L17}
\end{equation}%
Then for any $p_{0}$ and $\mathbf{p}_{\bot }$ the numbers $N_{n}^{\mathrm{cr}%
}$ are negligible if the range $\Omega _{3}$ is small enough,%
\begin{equation}
N_{n}^{\mathrm{cr}}\sim \sqrt{\xi _{1}^{2}-\lambda }\sqrt{\xi
_{2}^{2}-\lambda }\rightarrow 0\;\;\mathrm{if}\;\;\omega \rightarrow 0.
\label{L18}
\end{equation}

The $L$-constant field is one of a regularization for a constant uniform
electric field and it is suitable for imitating a small-gradient field. That
is reason why the $L$-constant field with a sufficiently large length $L$,
\begin{equation}
\sqrt{eE}L\gg \max \left\{ 1,m^{2}/eE\right\} ,  \label{L-large}
\end{equation}%
and $\omega \gg 1$ is of interest. In what follows, we suppose that these
conditions hold true and additionally assume that%
\begin{equation}
\sqrt{\lambda }<K_{\bot },  \label{L19}
\end{equation}%
where $K_{\bot }$ is any given number satisfying the condition$\;\sqrt{eE}%
L/2\gg K_{\bot }^{2}\gg \max \left\{ 1,m^{2}/eE\right\} .$

Let us analyze how the numbers $N_{n}^{\mathrm{cr}}$ depend on the
parameters $\xi _{1,2}$ and $\lambda $. Here we assume that $\chi =1$. Since
$N_{n}^{\mathrm{cr}}$ are even functions of $p_{0}$, we can consider only
the case of $p_{0}\leq 0$. In this case $\xi _{2}\geq \sqrt{eE}L/2$ is
large, $\xi _{2}\gg \max \left\{ 1,\lambda \right\} $, and the asymptotic
expansions of the WPCFs with respect to $\xi _{2}$ are valid. As for the
parameter $\xi _{1}$, the whole interval $-\sqrt{eE}L/2\leq \xi _{1}\leq -%
\sqrt{\lambda }$ can be divided in two parts:%
\begin{eqnarray}
&&(\mathrm{a})\;-\sqrt{eE}L/2\leq \xi _{1}\leq -K,  \notag \\
&&(\mathrm{b})\;-K<\xi _{1}\leq -\sqrt{\lambda },  \label{L20a}
\end{eqnarray}%
where $K$ is any given number satisfying the condition $\sqrt{eE}L/2\gg K\gg
K_{\bot }^{2}.$

In the case (a), using asymptotic expansions of the WPCFs given by Eq.~(\ref%
{a3}) in Appendix \ref{Ap2}, we obtain
\begin{eqnarray}
N_{n}^{\mathrm{cr}} &=&e^{-\pi \lambda }\left[ 1-\left( 1-e^{-\pi \lambda
}\right) ^{1/2}\frac{\sqrt{\lambda }}{2}\left( \frac{\sin \phi _{1}}{%
\left\vert \xi _{1}\right\vert ^{3}}+\frac{\sin \phi _{2}}{\left\vert \xi
_{2}\right\vert ^{3}}\right) +O\left( \left\vert \xi _{1}\right\vert
^{-4}\right) +O\left( \left\vert \xi _{2}\right\vert ^{-4}\right) \right] ,
\notag \\
\phi _{1,2} &=&\left\vert \xi _{1,2}\right\vert ^{2}-\lambda \ln \left(
\sqrt{2}\left\vert \xi _{1,2}\right\vert \right) +\arg \Gamma \left(
i\lambda /2\right) -\pi /4.  \label{L21}
\end{eqnarray}%
Consequently, the quantity (\ref{L21}) is almost constant over the wide
range of energy $p_{0}$ for any given $\lambda $ satisfying Eq.~(\ref{L19}).
One finds the same leading asymptotic term for scalar particles,%
\begin{equation}
N_{n}^{\mathrm{cr}}=e^{-\pi \lambda }\left[ 1+O\left( \left\vert \xi
_{1}\right\vert ^{-2}\right) +O\left( \left\vert \xi _{2}\right\vert
^{-2}\right) \right] .  \label{L21b}
\end{equation}

It should be noted that asymptotic forms of $N_{n}^{\mathrm{cr}}$ for
fermions and bosons were calculated in Ref. \cite{WongW88} only at $p_{0}=0$%
. Eqs.~(\ref{L21}) and (\ref{L21b}) contain these results as a particular
case if one restores the factor $2\sqrt{2}$ omitted for oscillating term in
the case of fermions in Ref. \cite{WongW88}. When $\sqrt{eE}L\rightarrow
\infty $, one obtains the well-known result for fermions and bosons in a
constant uniform electric field \cite{Nikis70a,Nikis79,Nikis70b},
\begin{equation}
N_{n}^{\mathrm{cr}}\rightarrow N_{n}^{\mathrm{uni}}=e^{-\pi \lambda },
\label{L22}
\end{equation}%
setting $K\rightarrow \infty $ in Eqs.~(\ref{L21}) and (\ref{L21b}),
respectively.

\begin{figure}[tbp]
\centering\includegraphics[scale=.6]{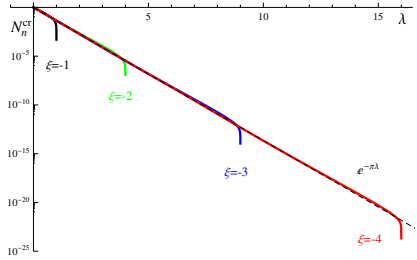}
\caption{The dependence on $\protect\lambda $ of $N_{n}^{\mathrm{cr}}$ for
different $\protect\xi =\protect\xi _{1}=-1,-2,-3,-4$.}
\label{2}
\end{figure}

In the range (b), using the only asymptotic expansions with respect to $\xi
_{2}$ given by Eq.~(\ref{a3}) in Appendix \ref{Ap2} and the exact form of $%
f_{2}^{(+)}(\xi _{1})$ given by Eq.~(\ref{L9}), we find
\begin{equation}
N_{n}^{\mathrm{cr}}=4e^{-\pi \lambda /4}\left[ \left( \sqrt{\xi
_{1}^{2}-\lambda }+\left\vert \xi _{1}\right\vert \right) \sqrt{\xi
_{1}^{2}-\lambda }+O\left( \left\vert \xi _{2}\right\vert ^{-3}\right) %
\right] ^{-1}\left\vert f_{2}^{(+)}(\xi _{1})\right\vert ^{-2}  \label{L23}
\end{equation}%
exactly with respect to $\xi _{1}$. The dependence on $\lambda <\xi _{1}^{2}$
of $N_{n}^{\mathrm{cr}}$ given by Eq.~(\ref{L23}) vs the function $e^{-\pi
\lambda }$ is found numerically for different $\xi _{1}=-1,-2,-3,-4$ and
presented in Fig.~\ref{2}. One can see that the asymptotic behavior of $%
e^{-\pi \lambda }$, which is typical for large $\left\vert \xi
_{1}\right\vert \gg \max \left\{ 1,\lambda \right\} $, appears if $\sqrt{\xi
_{1}^{2}-\lambda }\gtrsim 1$. We have $N_{n}^{\mathrm{cr}}\rightarrow 0$ as $%
\sqrt{\xi _{1}^{2}-\lambda }\rightarrow 0$ in accordance with the general
property (\ref{L15}). In particular, it is clear that the value $K=2$ is
sufficiently large for the case of small $\lambda \lesssim 1$. That is the
reason why we have retained explicitly oscillating terms in Eq.~(\ref{L21}).
We see that $N_{n}^{\mathrm{cr}}\lesssim e^{-\pi \lambda }$ in the range b).
The same conclusion can be made for bosons.

Let us consider the total number $N^{\mathrm{cr}}$ of pairs created by the $%
L $-constant field, which is defined by Eq.~(\ref{TN}) in Appendix \ref{Ap}.
Calculating this number in the fermionic case, one has to sum the
corresponding differential mean numbers $N_{n}^{\mathrm{cr}}$ over the spin
projections and over the transversal momenta $\mathbf{p}_{\bot }$\ and
energy $p_{0}$. Since the $N_{n}^{\mathrm{cr}}$ do not depend on the spin
polarization parameters $\sigma _{s}$, the sum over the spin projections
produces only the factor $J_{(d)}=2^{\left[ \frac{d}{2}\right] -1}$. The sum
over the momenta and the energy can be easily transformed into an integral
in the following way%
\begin{equation}
N^{\mathrm{cr}}=\sum_{\mathbf{p}_{\bot },p_{0}\in \Omega _{3}}\sum_{\sigma
}N_{n}^{\mathrm{cr}}=\frac{V_{\bot }TJ_{(d)}}{(2\pi )^{d-1}}\int_{\Omega
_{3}}dp_{0}d\mathbf{p}_{\bot }N_{n}^{\mathrm{cr}},  \label{L24}
\end{equation}%
where $V_{\bot }$ is the spatial volume of the ($d-1)$ dimensional
hypersurface orthogonal to the electric field direction and $T$ is the time
duration of the electric field. The total number of bosonic pairs created in
all possible states follows from Eq. (\ref{L24}) at $J_{(d)}=1$. Taking into
account that $N_{n}^{\mathrm{cr}}\lesssim e^{-\pi \lambda }$ in the range
(b), one obtains a rough estimation that the contribution from the range (b)
to the integral in the right-hand side of Eq. (\ref{L24}) is relatively
small,
\begin{equation*}
\int_{\left\vert \xi _{1}\right\vert <K}N_{n}^{\mathrm{cr}}dp_{0}\sim
e^{-\pi \lambda }\sqrt{eE}K\ .
\end{equation*}%
Therefore, the main contribution to integral (\ref{L24}) is due to an inner
subrange $D\subset \Omega _{3}$, which is defined by Eq.~(\ref{L19}) and to
the range (a) given by Eq.~(\ref{L20a}) for $p_{0}\leq 0$. Taking into
account that $N_{n}^{\mathrm{cr}}$ is an even function of $p_{0}$, we find
the complete subrange $D$ as%
\begin{eqnarray}
&&D:\sqrt{\lambda }<K_{\bot },\;\;\left\vert p_{0}\right\vert /\sqrt{eE}<%
\sqrt{eE}L/2-K,  \notag \\
&&\sqrt{eE}L/2\gg K\gg K_{\bot }^{2}\gg \max \left\{ 1,m^{2}/eE\right\} .
\label{L25}
\end{eqnarray}%
In this subrange $N_{n}^{\mathrm{cr}}\approx e^{-\pi \lambda }$ for both
fermions and bosons. Then one can find the total number of created particles
with given transversal momentum and spin polarization but with all possible
energies:
\begin{equation}
N_{\mathbf{p_{\perp }},\sigma }=\frac{T}{2\pi }\int_{\Omega
_{3}}dp_{0}N_{n}^{\mathrm{cr}}=\Delta e^{-\pi \lambda },\;\;\Delta =\frac{%
\sqrt{eE}T}{2\pi }\left[ \sqrt{eE}L+O(K)\right] .  \label{L26}
\end{equation}%
The factor $\Delta $ can be interpreted as a number of quantum states with a
given energy, in which the particles can be created. If $\sqrt{eE}L$ is big
enough, the dependence on $K$ and $K_{\bot }$ can be ignored, that is, the
form of $N_{n}^{\mathrm{cr}}$ is unchanged in the inner subrange $D$. Thus,
the definition of the subrange $D$ (\ref{L25}) can be also treated as the
stabilization condition for $N_{n}^{\mathrm{cr}}$.

Substituting Eq.~(\ref{L26}) into integral (\ref{L24}), performing the
integration over $\mathbf{p}_{\bot }$, and neglecting the exponentially
small contribution from the range $\sqrt{\lambda }\rightarrow K_{\bot }$, we
finally obtain
\begin{equation}
N^{\mathrm{cr}}=V_{\bot }Tn^{\mathrm{cr}},\;\;n^{\mathrm{cr}}=r^{\mathrm{cr}}%
\left[ L+\frac{O(K)}{\sqrt{eE}}\right] ,\;\;r^{\mathrm{cr}}=\frac{%
J_{(d)}\left( eE\right) ^{d/2}}{(2\pi )^{d-1}}\exp \left\{ -\pi \frac{m^{2}}{%
eE}\right\} .  \label{L27}
\end{equation}%
Here $n^{\mathrm{cr}}$ presents the total number density of pairs created
per unit time and unit surface orthogonal to the electric field direction on
an interval of the length $L$. The density $r^{\mathrm{cr}}=n^{\mathrm{cr}%
}/L $ is known in the theory of constant uniform electric field as the
pair-production rate (see the $d$ dimensional case in Ref.~\cite{GavG96a}).
Note that unlike the model of pair creation by the $T$-constant field \cite%
{GavG96a}, where $N^{\mathrm{cr}}$ is a function of the time duration of the
field, Eq.~(\ref{L27}) represents $N^{\mathrm{cr}}$ as a function of the
field length $L$. The $T$-constant and $L$-constant fields are physically
distinct; only in the asymptotic case, when $T\rightarrow \infty $ and $%
L\rightarrow \infty ,$ can one consider these fields as regularizations of a
constant uniform electric field given by two distinct gauge conditions on
the electromagnetic potentials $A^{\mu }\left( X\right) $.

Using expressions for the vacuum-to-vacuum transition probability [Eq. (\ref%
{cq31}) for fermions and a similar form in Ref.~\cite{GavGit15} for bosons],
we find
\begin{equation}
P_{v}=\exp \left( -\mu N^{\mathrm{cr}}\right) ,\;\;\mu =\sum_{j=0}^{\infty }%
\frac{(-1)^{(1-\kappa )j/2}}{(j+1)^{d/2}}\exp \left( -j\pi \frac{m^{2}}{eE}%
\right) \;,  \label{L28}
\end{equation}%
where $\kappa =+1$ for fermions and $\kappa =-1$ for bosons. In the
asymptotic case when $T\rightarrow \infty $ and $L\rightarrow \infty $, the
vacuum-to-vacuum transition probability (\ref{L28}) coincides with the
result obtained for the $T$-constant field \cite{GavG96a} and represents the
$d$ dimensional analog of the well-known Schwinger formula \cite{Schwi51}.

\section{Discussion\label{S4}}

\subsection{Length and time scales of small-gradient fields\label{SS4.1}}

It should be noted that the subrange $D\subset \Omega _{3}$, where a
stabilization condition for $N_{n}^{\mathrm{cr}}$ holds, is derived for an
arbitrary external field $E$ satisfying the uniform inequality (\ref{L-large}%
). We stress that the dimensionless parameter $\sqrt{eE}L$ plays an
important role in inequality (\ref{L-large}). Studying the particle creation
in the $L$-constant field, one can see that the stabilization of $N_{n}^{%
\mathrm{cr}}$ (\ref{L22}) for the energies $|p_{0}|<<eEL/2$ comes at
sufficiently large $L$ \footnote{%
We have restored $\hbar $ and $c$ here for convenience of the reader.},%
\begin{equation}
L\gg \Delta l_{0},\;\Delta l_{0}=\Delta l_{\mathrm{st}}\max \left\{
1,\lambda \right\} ,\;\Delta l_{\mathrm{st}}=\left( \hbar c/eE\right) ^{1/2}.
\label{L29a}
\end{equation}%
The characteristic length $\Delta l_{0}$ can be called the stabilization
length for a given $\lambda $. We note that in the latter equation there
appears an uniform length scale $\Delta l_{\mathrm{st}}$. In addition, we
can define another specific uniform length scale $\Delta l_{\mathrm{st}%
}^{m}, $%
\begin{equation}
\Delta l_{\mathrm{st}}^{m}=\Delta l_{\mathrm{st}}\max \left\{
1,c^{3}m^{2}/\hslash eE\right\} ,  \label{L29}
\end{equation}%
which appears under the stabilization condition. The length scale $\Delta l_{%
\mathrm{st}}^{m}$ plays the role of the characteristic length to form the
distribution (\ref{L22}) for all $\lambda $ from the subrange $D$. The
latter distribution is typical for a uniform ($L\rightarrow \infty $)
electric field. Note that the exponential form~(\ref{L22}) plays the role of
a cutoff factor over $p_{\bot }$ such that only contributions from
relatively small transversal momenta, $p_{\bot }^{2}/eE\lesssim $ $\max
\left\{ 1,m^{2}/eE\right\} ,$ are essential. The length $\Delta l_{\mathrm{st%
}}^{m}$ can be called the uniform stabilization length. Note that the
characteristic value $m^{2}/eE$ can be represented as the ratio of two
characteristic lengths: $c^{3}m^{2}/\hslash eE=\left( \Delta l_{\mathrm{st}%
}/\Lambda _{C}\right) ^{2}$, where $\Lambda _{C}=\hbar /mc$ is the Compton
wavelength. We are primarily interested in strong electric fields, $\left(
\Delta l_{\mathrm{st}}/\Lambda _{C}\right) ^{2}\lesssim 1$. In this case,
inequality (\ref{L-large}) is simplified to the form $L/\Delta l_{\mathrm{st}%
}\gg 1$, in which the Compton wavelength is absent. We see that the scale $%
\Delta l_{\mathrm{st}}$ plays the role of the uniform stabilization length
for a strong electric field. This means that such a strong field has a
macroscopic length for the problem under consideration, and that $\Delta l_{%
\mathrm{st}}$ is the characteristic length that differentiates between
fields that have microscopical or macroscopic inhomogeneity; i.e., it plays
the role that the Compton wavelength plays in the case of a weak field.

We assume that our constant external electric field exists during a
macroscopically large time period $T$, which means that $T$ is sufficiently
large compared to the time scale $\Delta t_{\mathrm{st}}^{m}=\Delta l_{%
\mathrm{st}}^{m}/c$, $\Delta t_{\mathrm{st}}^{m}\ll T$. On the other hand,
the pair creation is transient process and the applicability of the constant
field approximation is limited by the smallness of the backreaction. For
example, in $d=3+1$ and $L/\Delta l_{\mathrm{st}}\gg 1$, depletion of an
electric field due to the backreaction implies a restriction%
\begin{equation}
1\ll \left( cT/\Delta l_{\mathrm{st}}\right) ^{2}\ll \frac{\pi ^{2}}{J\alpha
}\,\exp \left( \pi \frac{c^{3}m^{2}}{\hslash eE}\right) ,  \label{L30}
\end{equation}%
on time $T$ for a given electric field strength. Here $\alpha $ is the fine
structure constant and $J$ is the number of the spin degrees of freedom ($%
J=1 $ for scalar particles, $J=2$ for spin-$1/2$ particles, and $J=3$ for
vector particles), see \cite{GavG08}. Thus, there is a {window} in the
parameter range of $E$ and $T$ where the approximation of the constant
external field is consistent. For QCD with a constant $SU(3)$ chromoelectric
field $E^{a}$ ($a=1,\ldots ,8$) (during the period when the produced partons
can be treated as weakly coupled due to the property of asymptotic freedom
in QCD), and at low temperatures $\theta \ll q\sqrt{C_{1}}T,$ the
consistency restriction for the dimensionless parameter $q\sqrt{C_{1}}T^{2}$
has the form%
\begin{equation}
1\ll q\sqrt{C_{1}}T^{2}\ll \pi ^{2}/3q^{2}\,,  \label{L31}
\end{equation}%
where $q$ is the coupling constant and $C_{1}=E^{a}E^{a}$ is a Casimir
invariant for{\large \ }$SU(3)$.

In order to estimate the boundary effects of quasiuniform electric fields,
one can study another example of a small-gradient field. We take as our
example the Sauter field \cite{Sauter-pot},
\begin{equation*}
E(x)=E\cosh ^{-2}\left( x/L_{\mathrm{S}}\right) ,\ \ L_{\mathrm{S}}>0,
\end{equation*}%
which can be considered as another regularization for the constant uniform
electric field when $eEL_{\mathrm{S}}^{2}\gg 1$. It can be shown (see, e.g.,
Ref.~\cite{GavGit15}) that in the wide range of energies where $\left\vert
p_{0}\right\vert \ll eEL_{\mathrm{S}}$ and
\begin{equation}
L_{\mathrm{S}}\gg \Delta l_{\mathrm{f}},\ \ \Delta l_{\mathrm{f}}=\Delta l_{%
\mathrm{st}}\max \left\{ 1,\sqrt{\lambda }\right\} ,  \label{L32}
\end{equation}%
the numbers $N_{n}^{\mathrm{cr}}$ do not, in fact, depend on the parameter $%
L_{\mathrm{S}}$; they have the form (\ref{L22}), which coincides with the
differential number of created particles in a uniform electric field \cite%
{Nikis79,Nikis70b}. For large $L_{\mathrm{S}}$ the Sauter field varies
slowly and nearly coincides with the uniform field on the distance $|x|<L_{%
\mathrm{S}}$. Then, $\Delta l_{\mathrm{f}}$ is a characteristic length of
the stabilization for a given $\lambda $ in this field. For any given $%
\lambda >1$ the stabilization length of the Sauter field, $\Delta l_{\mathrm{%
f}}$, is less than the stabilization length of the $L$-constant field, $%
\Delta l_{0}$. In the case of a strong field when there exists $\lambda <1$,
both stabilization lengths have the same value, $\Delta l_{\mathrm{st}}$. We
conclude that the stabilization process for $\lambda >1$ depends on the
boundary effects. In addition, one can see that a smooth asymptotic decrease
of the Sauter field at $|x|\gtrsim L_{\mathrm{S}}$ affects the quantum
system less than a sharp disappearance of the $L$-constant field at $|x|=L/2$%
. Thus, a stabilization length for a given $\lambda >1$ is not a universal
characteristic; it depends on the field form. Since the Sauter field for $%
\lambda >1$ is nearly uniform on the interval $\sim \Delta l_{\mathrm{f}}$
and is strong enough for the stabilization there, one can interpret this
interval as a universal length of pair formation, which does not depend on
the field behavior in sufficiently remote regions. One can extrapolate this
interpretation of $\Delta l_{\mathrm{f}}$ for any field. A semiclassical
consideration is in agreement with this interpretation.{\huge \ }Thus, the
uniform electric field produces a work $eE\Delta l_{\mathrm{f}}=\pi _{\bot }$
acting on a charge $e$ on the distance $\Delta l_{\mathrm{f}}$ , such that a
virtual electron-positron pair obtains the energy $2\pi _{\bot }$, this is
the sum of the kinetic and the rest energy, and can be materialized.

One may ask a question: For which maximal potential differences some of
total effects of particle creation are the same by the Sauter field and by
the $L$-constant field? The potential differences are$\mathbb{\ }2eEL_{%
\mathrm{S}}$ and $eEL$, respectively, for these fields.\ For example, by
comparing the number of states with the given energy in which particles can
be created, one sees that at any finite $\lambda $ the effect of the Sauter
field at large $L_{\mathrm{S}}$ \cite{GavGit15} is equivalent to the effect
of the $L$-constant field at the large $L$, given by Eq.~(\ref{L26}), with
the identification $L_{\mathrm{S}}=\sqrt{\lambda }L$. Performing the
integration over $\mathbf{p}_{\bot }$, one can compare the results for the
total number of created particles $N^{\mathrm{cr}}$ for both cases and can
see the equivalence of these total numbers with the identification $L_{%
\mathrm{S}}=L/\delta $, where

\begin{equation*}
\delta =\sqrt{\pi }\Psi \left( \frac{1}{2},-\frac{d-2}{2};\pi \frac{m^{2}}{eE%
}\right)
\end{equation*}%
and $\Psi \left( a,b;x\right) $ is the confluent hypergeometric function
\cite{BatE53}. If the field is weak (i.e., $m^{2}/eE\gg $ $1$) using the
asymptotic expression for the $\Psi $-function, one obtains that $\delta
\approx \sqrt{eE}/m$. This can be treated as the above identification $L_{%
\mathrm{S}}=\sqrt{\lambda }L$ at $p_{\bot }\rightarrow 0$, and means that
only small $p_{\bot }\rightarrow 0$ are essential. In the case of a very
strong field, $m^{2}/eE\ll 1$, one obtains from the Ref.~\cite{BatE53} that
the leading term for $\delta $ does not depend on the parameter $m^{2}/eE$,
\begin{equation}
\delta \approx \sqrt{\pi }\Gamma \left( d/2\right) /\Gamma \left(
d/2+1/2\right) .  \label{L33}
\end{equation}%
For example, $\delta \approx \pi /2$ if $d=3$ and $\delta \approx 4/3$ if $%
d=4$.

Another total quantity is the vacuum-to-vacuum transition probability $P_{v}$%
. The probability $P_{v}$ obtained for the Sauter field reads \cite{GavGit15}
\begin{eqnarray}
&&P_{v}=\exp \left( -\mu ^{\mathrm{S}}N^{\mathrm{cr}}\right) ,\;\;\mu ^{%
\mathrm{S}}=\sum_{j=0}^{\infty }\frac{(-1)^{(1-\kappa )j/2}\epsilon _{j+1}}{%
(j+1)^{d/2}}\exp \left( -j\pi \frac{m^{2}}{eE}\right) \;,  \notag \\
&&\epsilon _{j}=\delta ^{-1}\sqrt{\pi }\Psi \left( \frac{1}{2},-\frac{d-2}{2}%
;j\pi \frac{m^{2}}{eE}\right) .  \label{L34}
\end{eqnarray}%
Comparing the probability (\ref{L34}) and (\ref{L28}), obtained for the $L$%
-constant field, one can establish relation between parameters $L_{\mathrm{S}%
}$ and $L$. If the field is weak, $m^{2}/eE\gg $ $1$, then $\;\epsilon
_{j}\approx j^{-1/2}$ and $\mu ^{\mathrm{S}}\approx \mu \approx 1$, and the
identification $L_{\mathrm{S}}=L/\delta \approx Lm/\sqrt{eE}$ is the same as
the one extracted from the comparison of total numbers $N^{\mathrm{cr}}$. In
the case of a strong field, all the terms with different $\epsilon _{j}$
contribute significantly to the sum in Eq.~(\ref{L34}) if $j\pi m^{2}/eE\sim
1$, and the expression for{\huge \ }$P_{v}$ in (\ref{L34}) differs
essentially from the one in (\ref{L28}). However, for the very strong field
if $j\pi m^{2}/eE\ll 1$, the leading contribution of $\epsilon _{j}$ has a
quite simple form $\epsilon _{j}\approx 1$. In this case%
\begin{equation*}
\mu ^{\mathrm{S}}\approx \mu \approx \sum_{j=0}^{\infty }\frac{%
(-1)^{(1-\kappa )j/2}}{(j+1)^{d/2}}
\end{equation*}%
and the identification $L_{\mathrm{S}}=L/\delta $ is the same as the one
extracted from the comparison of the total numbers $N^{\mathrm{cr}}$, where $%
\delta $ is given by Eq.~(\ref{L33}).

It should be noted that total contributions to vacuum mean values, e.g., to
the mean electric current and the mean energy-momentum tensor, are usually
of interest in small-gradient fields. These quantum effects are proportional
to corresponding sums of differential numbers of created particles, e.g., to
the number of particles with given transversal momentum and to the total
number of created particles. Consequently, it is useful to derive a relation
between these total numbers and parameters $L_{\mathrm{S}}$\ and $L$. Such a
relation derived from the vacuum-to-vacuum transition probability $P_{v}$\
is interesting for semiclassical approaches based on Schwinger's technique
\cite{Schwi51}. For the weak field, $m^{2}/eE\gg $\ $1$, the identification $%
L_{\mathrm{S}}\approx Lm/\sqrt{eE}$\ follows from the comparison of $N^{%
\mathrm{cr}}$\ and $P_{v}$\ in $L$\ -constant and Sauter fields.

The above consideration allows us to conclude that for both Sauter and $L$%
-constant fields the differential and total effects of pair creation in
sufficiently large regions (where the fields are nearly uniform) and for
finite energies of particles, are not significantly affected by the field
behavior far away from the region. Extrapolating these results, one may
believe that in any quasiuniform electric field $\approx E$ on the region $%
L\gg \Delta l_{\mathrm{st}}^{m}$, that vanishes out the region,
particle-creation effects must not depend on the details of the switching
off. Therefore, calculations in an $L$-constant field are quite
representative for a large class of small-gradient electric fields.

It is well known that at certain conditions (the so-called charge neutrality
point) electronic excitations in graphene monolayer behave as relativistic
Dirac massless fermions in $2+1$ dimensions, with the Fermi velocity $%
v_{F}\simeq 10^{6}$ \textrm{m/s} playing the role of the speed of light in
relativistic particle dynamics; see details in recent reviews \cite%
{dassarma,VafVish14}. Then in the range of applicability of the Dirac model
to the graphene physics, any electric field is strong. There appears a
length scale specific to graphene (and to similar nanostructures with the
Dirac fermions),
\begin{equation}
\Delta l_{\mathrm{st}}^{g}=\left( \hbar v_{F}/eE\right) ^{1/2},  \label{L30g}
\end{equation}%
which plays the role of the stabilization length. The generation of a mass
gap in the graphene band structure is an important fundamental and practical
problem under current research, see, e.g., the recent report \cite{gap15} on
the fabrication of a large band gap, $0.5$ eV, in epitaxially grown graphene
samples. In the presence of the mass gap $\Delta \varepsilon =mv_{F}^{2}$,
the stabilization condition has general form (\ref{L-large}) that involves a
length scale $\Delta l_{\mathrm{st}}^{gm}=\Delta l_{\mathrm{st}}^{g}\max
\left\{ 1,(\Delta \varepsilon )^{2}/\hslash v_{F}eE\right\} $. In this case
the strong field condition reads $(\Delta \varepsilon )^{2}/v_{F}\hbar eE\ll
1$. It is shown in \cite{lewkowicz-10a,lewkowicz-10b} that the time scale $%
\Delta t_{\mathrm{st}}^{g}=\Delta l_{\mathrm{st}}^{g}/v_{F}$ appears for the
tight-binding model as the time scale when the perturbation theory with
respect to electric field breaks down ($\Delta t_{\mathrm{st}}^{g}\gg
t_{\gamma }$, where the microscopic time scale is $t_{\gamma }=\hbar /\gamma
\simeq 0.24\;\mathrm{fs}$, with $\gamma =2.7$ eV being the hopping energy),
and the \textrm{dc} response changes from the linear-in-$E$
duration-independent\emph{\ } regime to a nonlinear-in-$E$ and
duration-dependent regime.\ The length between two electrodes $L$ is less
than the length of a graphene flake $L^{g}$, $L<L^{g}$. In the experimental
situation described in Ref. \cite{Van+etal10}, a constant voltage between
two electrodes connected to the graphene was applied, and current-voltage
characteristics ($I-V$) are measured within exposition time $T_{ex}\sim 1$
\textrm{s}, which is a very large time scale compared with the ballistic
flight time $T_{bal}=L^{g}/v_{F}$ (the time that the electron spends to
cross the length $L^{g}$). The time dimension $T$ is macroscopically large, $%
\Delta t_{\mathrm{st}}^{g}\ll T$, and less than the time $T_{ex}$. The
external constant electric field can be considered as a good approximation
of the effective mean field as long as the field produced by the induced
current of created particles is negligible compared to the applied field. {%
This gives the consistency restriction }$T\ll \Delta t_{\mathrm{br}}=\Delta
t_{\mathrm{st}}^{g}\pi /4\alpha $ \cite{GavGitY12}, where $\alpha $ is the
fine structure constant. Thus, there is a {window} in the parameter range of
$E$ and $T$ where the model with constant external field is consistent,
\begin{equation}
\Delta t_{\mathrm{st}}^{g}\ll T\ll \Delta t_{\mathrm{br}}.  \label{L31g}
\end{equation}

For example, let us assume that{\ $T=T_{bal}$}. In typical experiments, $%
L^{g}\sim 1\;\mathrm{\mu m}$, so that $T_{bal}\sim 10^{-12}\;\mathrm{s}$.
Then, we obtain from Eq.~(\ref{L31g}) the following restrictions on the
external electric field:
\begin{equation*}
7\times 10^{2}\;\mathrm{V/m}\ll E\ll 8\times 10^{6}\;\mathrm{V/m}\,.
\end{equation*}%
Since the voltage is $V=EL$ and assuming that $L\approx L^{g}$ one finds the
inequalities
\begin{equation*}
7\times 10^{-4}\,\mathrm{V}\ll V\ll 8\,\mathrm{V}\,.
\end{equation*}%
These voltages are in the range typically used in experiments with graphene.
In this electric field, we find that the range of length scale $\Delta l_{%
\mathrm{st}}^{g}$ satisfies the inequalities
\begin{equation*}
0.01\;\mathrm{\mu m}\ll \Delta l_{\mathrm{st}}^{g}\ll 1\;\mathrm{\mu m.}
\end{equation*}%
This shows that QED with an $L$-constant field is a good model to describe
the quantum effects in graphene placed in a constant external electric field.

\subsection{Connection between the vacuum instability in external
electromagnetic and gravitational fields\label{SS4.2}}

To gain  insight into universal features of particle creation from vacuum,
it is useful to compare effects caused by external fields of different
nature. The situation with a uniform electric field confined between two
capacitor plates has many similarities with both the chromoelectric flux
tube and the de Sitter case, see, e.g.,  Refs.~\cite%
{GelTan15,GavGT06,AndMot14,AkhmP15} for the review. The idea that particle
creation by an electric field has similarity with particle emission from
black holes calculated by Hawking \cite{Haw75} was tested for the first time
by Frolov and Gitman in Refs. \cite{b24}. Since the Hawking radiation was
considered to be a component of created particles (particles out of the
horizon), the authors of the latter works derived for their comparison a
reduced density matrix of electrons created by a quasiconstant electric
field. Using the equivalence principle, they obtained almost Hawking
distribution (up to a factor of 2 away from the Hawking temperature). Then,
taking vacuum polarization effects into account, we showed that the
distribution of electrons created by a slowly varying uniform electric field
-- in particular, by the $T$-constant field -- can be written in a general
Hawking-like thermal form, in which the Hawking temperature is reproduced
exactly \cite{GavG96a}. One can establish a similar connection for the case
of strong electrostatic inhomogeneous fields with of a small gradient. To do
this, we will use the model of the $L$-constant field.

Note that the $T$-constant and $L$-constant electric fields produce the same
quantum effects (coinciding with ones caused by a constant uniform electric
field) in the limiting case, $T\rightarrow \infty $ and $L\rightarrow \infty
$, if these limits exist (see discussions of the applicability of the model
of a constant uniform electric field in Refs.~\cite{GG06-08,GavG08,GavGitY12}%
). However, the $T$-constant and $L$-constant fields describe different
physical situations in the general case. This is the reason why we cannot
follow the method used in Ref.~\cite{GavG96a} \textrm{\ }to study the
consequences of the equivalence principle.

The phenomenon of particle emission from black holes was first considered by
Hawking \cite{Haw75}, who  calculated the mean numbers $N_{n}$ of particles
created by static gravitational field of a black hole in a specific thermal
environment,
\begin{equation}
N_{n}=\left\{ \exp \left[ 2\pi \frac{\omega }{g_{(H)}}\right] +\kappa
\right\} ^{-1}.  \label{ec1}
\end{equation}%
Here $\omega $ is the energy of a created particle, which we suppose to be
dependent on quantum numbers $n$, $g_{(H)}=GM/r_{g}^{2}$, where $r_{g}$ is
the gravitational radius of mass $M$, so that $g_{(H)}$ is free-falling
acceleration at this radius. This spectrum was interpreted as a Planck
distribution with the temperature $\theta _{(H)}=g_{(H)}\left( 2\pi
k_{B}\right) ^{-1}$ ($k_{B}$ is the Boltzmann constant). As before $\kappa
=+1$ for fermions and $\kappa =-1$ for bosons. It is also known \cite%
{Unruh76} that an observer with a constant acceleration $g_{(R)}$ (with
respect to its proper time) will register some particles (Rindler particles)
in the Minkowski vacuum. The mean numbers of Rindler bosons have the same
Planck form as (\ref{ec1}) (with $\kappa =-1$), where one has to replace $%
g_{(H)}$ by $g_{(R)}$, so that the corresponding temperature is $\theta
_{(R)}=g_{(R)}\left( 2\pi k_{B}\right) ^{-1}$. One can find many other
examples when particle creation in external gravitation fields (and due to a
nontrivial topology) can be described by means of an effective temperature;
see Refs.~\cite{BirDav82,GMM94,Pad05,NJBN12} for reviews.

The distribution (\ref{L22}) obtained for the $L$-constant field does not
have a thermal form at first blush. Nevertheless, in the framework of a
semiclassical consideration and using some results of the present paper one
can find a close connection to a thermal-like form. As established, both
electrons and positrons with given transversal momentum $\mathbf{p}_{\perp }$
and zero longitudinal momentum are created in a subregion of the region $S_{%
\mathrm{int}}$ with the kinetic and rest energy $\pi _{\bot }$ per particle.
At the same time, the created electrons and positrons are accelerated by the
electric field along the axis $x$ to the left and to the right,
respectively. Note that in the subrange $D\subset \Omega _{3}$, where a
stabilization condition for $N_{n}^{\mathrm{cr}}$ holds, the width $\Delta
l_{\mathrm{f}}$ of the pair formation subregion is small compared to the
distance $L$. Finally, the particles appear on the left and the right of the
step already having  ultrarelativistic velocities and longitudinal momenta $%
\left\vert p^{\mathrm{L}}\right\vert $ and $\left\vert p^{\mathrm{R}%
}\right\vert $, respectively, given by Eq.~(\ref{L3}). Using classical
equations of motion in a constant uniform electric field $d\mathbf{P}/dt=e%
\mathbf{E}$, one finds final accelerations for both kinds of particles as
\begin{equation}
g\left( \mathrm{L/R}\right) =eE/\left\vert \pi _{0}\left( \mathrm{L/R}%
\right) \right\vert \approx 2/L,  \label{L35}
\end{equation}%
respectively.

We can improve the classical consideration by taking into account properties
of the physical vacuum in the $L$-constant field. Within the general
construction ~\cite{GavGit15}, it is assumed \ that electrons and positrons
in one of corresponding asymptotic regions, $S_{\mathrm{L}}$ and $S_{\mathrm{%
R}}$, occupy quasistationary states; i.e., they are described by wave
packets that maintain their form on a sufficiently large distance in one of
the corresponding asymptotic regions.\ Such wave packets are superpositions
of the corresponding plane waves from some subrange of energies $%
p_{0}\subset D$. In the semiclassical approximation, we assume that the
energy of a particle $\bar{p}_{0}$ is an average value of these energies $%
p_{0}\subset D$. Then the total energy of a pair created with given $\bar{p}%
_{0}$ , $\mathbf{p}_{\perp }$ is a sum $\bar{\pi}_{0}\left( \mathrm{L}%
\right) +\left\vert \bar{\pi}_{0}\left( \mathrm{R}\right) \right\vert $,
where $\bar{\pi}_{0}\left( \mathrm{L/R}\right) =\bar{p}_{0}-U_{\mathrm{L/R}}$%
, i.e., this energy is equal to the total field work $\bar{\pi}_{0}\left(
\mathrm{L}\right) +\left\vert \bar{\pi}_{0}\left( \mathrm{R}\right)
\right\vert =\mathbb{U}$. We note that this field work was partially used
for the pair creation and partially for their further acceleration, such
that before leaving the region $S_{\mathrm{int}}$ they have gained the
following average longitudinal momenta%
\begin{equation*}
\left\vert \bar{p}^{\mathrm{L/R}}\right\vert =\sqrt{\left[ \bar{\pi}%
_{0}\left( \mathrm{L/R}\right) \right] ^{2}-\pi _{\bot }^{2}}.
\end{equation*}%
At the same time the corresponding part of\textrm{\ }energy was lost for the
field restricted in the region $S_{\mathrm{int}}$. Let us estimate this
energy considering wave packets $^{+}\psi _{x_{\mathrm{F}}}\left( X\right) $
and $_{+}\psi _{x_{F}}\left( X\right) $, which consist of partial waves of
an $\mathrm{out}$-electron $^{+}\psi _{n}\left( X\right) $ and an $\mathrm{%
out}$-positron $_{+}\psi _{n}\left( X\right) $, and which have focal planes $%
x=x_{\mathrm{F}}$ somewhere in $S_{\mathrm{L}}$ and $S_{\mathrm{R}}$,
respectively. One can construct such wave packets using a procedure
described in Appendix D of Ref.~\cite{GavGit15}. The energy flux of the
field $\psi _{x_{\mathrm{F}}}$ through a surface $x=x^{\mathrm{out}}$ is
defined as
\begin{equation*}
F\left( x\right) =\int_{x=x^{\mathrm{out}}}\psi _{x_{\mathrm{F}}}^{\dag
}\left( X\right) \hat{p}_{x}\psi _{x_{\mathrm{F}}}\left( X\right) d\mathbf{r}%
_{\bot }\ .
\end{equation*}%
Then through any plane $x=x_{\mathrm{R}}^{\mathrm{out}}\in S_{\mathrm{R}}$ a
positron carries away the energy $\left\vert \bar{p}^{\mathrm{R}}\right\vert
$ for the time $T$ , whereas through any plane $x=x_{\mathrm{L}}^{\mathrm{out%
}}\in S_{\mathrm{L}}$ an electron carries away the energy $\left\vert \bar{p}%
^{\mathrm{L}}\right\vert $ for the time $T$ . Thus, the field work spent for
a pair acceleration is $\left\vert \bar{p}^{\mathrm{L}}\right\vert
+\left\vert \bar{p}^{\mathrm{R}}\right\vert $. Consequently, the work spent
for the creation of a pair in a given state is

\begin{equation}
2\omega =\mathbb{U-}\left\vert \bar{p}^{\mathrm{L}}\right\vert -\left\vert
\bar{p}^{\mathrm{R}}\right\vert \approx \lambda g,\ \ g=\left[ g\left(
\mathrm{L}\right) +g\left( \mathrm{R}\right) \right] /2,  \label{L36}
\end{equation}%
where $\omega $ is the work spent for the creation of a particle from a
pair, $\lambda $ is given by Eq.~(\ref{L5}), and $g\left( \mathrm{L/R}%
\right) $ are given by Eq.~(\ref{L35}).

Then the distribution (\ref{L22}) can be rewritten in the following form
\begin{equation}
N_{n}^{\mathrm{uni}}=\exp \left\{ -2\pi \frac{\omega }{g}\right\} .
\label{L37}
\end{equation}%
The energy of a particle in the Hawking formula (\ref{ec1}) can be treated
as the work, $\omega $, that a gravitational field has spent for the
creation of a particle. Then Eq.~(\ref{L37}) is, in fact, the Boltzmann
distribution with the temperature $\theta =g\left( 2\pi k_{B}\right) ^{-1}$
having literally the Hawking form. Thus, once again we see that the
distributions of particles created by electromagnetic and gravitational
fields have similar thermal structures.

It is a direct consequence of the equivalence principle that the effective
temperature $\theta $ of distribution (\ref{L37}) has literally the Hawking
form. Regarding the distinction between the Planck and the Boltzmann
distributions, we believe that the Planck distribution for the Hawking case
necessarily arises due to the appearance of an event horizon (there is a
boundary of the domain of the Hamiltonian); that is, it is due to the
condition for which the space domains of particle and antiparticle vacua are
not the same. In contrast to this, in an electric field, we deal with both
the particle vacuum and the antiparticle vacuum defined over the entire
space, see Ref.~\cite{GavGit15}; that is why these space domains coincide.

\section{Conclusion\label{S5}}

In this Conclusion we would like to try to characterize the place of the
present article among the numerous works devoted to the effect of pair
creation from a vacuum by an external electromagnetic field. First, we
recall that in the area under discussion the possibility of obtaining any
nonperturbative result is, as a rule, based on the existence of special
exact solutions of the Dirac equation with external electromagnetic fields
that are able to create pairs from the vacuum. However, it is well known
that there exist few such field configurations and corresponding exact
solutions. We believe that, among these, the $T$-constant and the $L$%
-constant electric fields with sufficiently large parameters $T$\ and $L$\
have \emph{a priority}, because studying the effect in such relatively
simple field configurations allows one to understand typical physical
characteristics of the effect in wide classes of external fields. The study
of a pair creation effect{\large \ }in the $T$-constant electric fields
already has a long story, which we{\large \ }cited in the Introduction. Here
almost all local and global characteristics of the effect were calculated in
detail using the well-developed formulation of QED with $t$-electric
potential steps. A similar study for $L$-constant electric fields\ did not
exist until the present, as a formulation of QED with $x$-electric potential
steps, which is sufficient for this problem, was developed by us a short
time ago in Ref. \cite{GavGit15}. Our present article, then, contains for
the first time a consistent QED treatment of the pair creation effect in the
$L$-constant electric field, a treatment that is free from misunderstandings
of a naive one-particle consideration. Moreover, the presented sets of
stationary solutions as well as their interpretation, probabilities, vacuum
mean values, and analysis of length and time scales, constitute a possible
basis for future research in strong-field QFT of small-gradient fields.

\subparagraph{\protect\large Acknowledgement}

S. P. G. and D..M. G. were supported by a grant from the Russian Science
Foundation, Research Project No. 15-12-10009.

\appendix

\section{Basic elements of QED with $x$-electric critical potential steps
\label{Ap}}

In this appendix, we briefly present  some basic constructions of QED with
an $L$-constant electric field that follow from the general formulation of
QED with $x$-electric potential step \cite{GavGit15} and the results of
Sect. \ref{S2}.

Solutions of the Dirac equation, $_{\zeta }\psi _{n}\left( X\right) $ and $%
^{\zeta }\psi _{n}\left( X\right) $, can be subjected to the following
orthonormality conditions on the $x=\mathrm{const}$ hyperplane: \emph{\ }%
\begin{eqnarray}
&&\left( \ _{\zeta }\psi _{n},\ _{\zeta ^{\prime }}\psi _{n^{\prime
}}\right) _{x}=\zeta \eta _{\mathrm{L}}\delta _{\zeta ,\zeta ^{\prime
}}\delta _{n,n^{\prime }},\ \ \eta _{\mathrm{L}}=\mathrm{sgn\ }\pi
_{0}\left( \mathrm{L}\right) ,  \notag \\
&&\left( \ ^{\zeta }\psi _{n},\ ^{\zeta ^{\prime }}\psi _{n^{\prime
}}\right) _{x}=\zeta \eta _{\mathrm{R}}\delta _{\zeta ,\zeta ^{\prime
}}\delta _{n,n^{\prime }},\ \ \eta _{\mathrm{R}}=\mathrm{sgn\ }\pi
_{0}\left( \mathrm{R}\right) \ ;  \notag \\
&&\left( \psi ,\psi ^{\prime }\right) _{x}=\int \psi ^{\dag }\left( X\right)
\gamma ^{0}\gamma ^{1}\psi ^{\prime }\left( X\right) dtd\mathbf{r}_{\bot }\ .
\label{c3}
\end{eqnarray}%
We consider our theory in a large spacetime box that has a spatial volume $%
V_{\bot }=\prod\limits_{j=2}^{D}K_{j}$ and the time dimension $T$, where all
$K_{j}$ and $T$ are macroscopically large. The integration over the
transverse coordinates is fulfilled from $-K_{j}/2$ to $+K_{j}/2$, and over
the time $t$ from $-T/2$ to $+T/2$. The limits $K_{j}\rightarrow \infty $
and $T\rightarrow \infty $ are assumed in final expressions.

The time-independent inner product for any pair of solutions of the Dirac
equation, $\psi _{n}\left( X\right) $ and $\psi _{n^{\prime }}^{\prime
}\left( X\right) $, is defined on the $t=$\textrm{const} hyperplane as
follows:
\begin{equation}
\left( \psi _{n},\psi _{n^{\prime }}^{\prime }\right) =\int_{V_{\bot }}d%
\mathbf{r}_{\bot }\int\limits_{-K^{\left( \mathrm{L}\right) }}^{K^{\left(
\mathrm{R}\right) }}\psi _{n}^{\dag }\left( X\right) \psi _{n^{\prime
}}^{\prime }\left( X\right) dx,\ \   \label{t4}
\end{equation}%
where the improper integral over $x$ in the right-hand side of Eq. (\ref{t4}%
) is reduced to its special principal value to provide a certain additional
property important for us and the limits $K^{\left( \mathrm{L}/\mathrm{R}%
\right) }\rightarrow \infty $ are assumed in final expressions. The
following orthonormality relations are on the $t=$\textrm{const} hyperplane
\begin{eqnarray}
&&\left( \ _{\zeta }\psi _{n},\ _{\zeta }\psi _{n^{\prime }}\right) =\left(
\ ^{\zeta }\psi _{n},\ ^{\zeta }\psi _{n^{\prime }}\right) =\delta
_{n,n^{\prime }}\mathcal{M}_{n}\ ,\ \ n\in \Omega _{1}\cup \Omega _{3}\cup
\Omega _{5}\ ;  \notag \\
&&\left( \psi _{n},\psi _{n^{\prime }}\right) =\delta _{n,n^{\prime }}%
\mathcal{M}_{n},\;n\in \Omega _{2}\cup \Omega _{4}\ ,  \notag \\
&&\mathcal{M}_{n}=2\frac{K^{\left( \mathrm{R}\right) }}{T}\left\vert \frac{%
\pi _{0}\left( \mathrm{R}\right) }{p^{\mathrm{R}}}\right\vert \left\vert
g\left( _{+}\left\vert ^{+}\right. \right) \right\vert ^{2},\ \ n\in \Omega
_{1}\cup \Omega _{5},  \notag \\
&&\mathcal{M}_{n}=2\frac{K^{\left( \mathrm{R}\right) }}{T}\left\vert \frac{%
\pi _{0}\left( \mathrm{R}\right) }{p^{\mathrm{R}}}\right\vert \left\vert
g\left( _{+}\left\vert ^{-}\right. \right) \right\vert ^{2},\ \ n\in \Omega
_{3},  \notag \\
&&\mathcal{M}_{n}=2\frac{K^{\left( \mathrm{L}\right) }}{T}\left\vert \frac{%
\pi _{0}\left( \mathrm{L}\right) }{p^{\mathrm{L}}}\right\vert ,\ \ n\in
\Omega _{2},\mathrm{\;}\mathcal{M}_{n}=2\frac{K^{\left( \mathrm{R}\right) }}{%
T}\left\vert \frac{\pi _{0}\left( \mathrm{R}\right) }{p^{\mathrm{R}}}%
\right\vert ,\ \ n\in \Omega _{4}.  \label{i12}
\end{eqnarray}%
All the wave functions having different quantum numbers $n$ are orthogonal,
and
\begin{eqnarray}
&&\left( \ _{\zeta }\psi _{n},\ ^{-\zeta }\psi _{n}\right) =0,\ \ n\in
\Omega _{1}\cup \Omega _{5}\ ,\ _{\ \zeta }\psi _{n}\ \mathrm{and}\ \
^{-\zeta }\psi _{n}\ \mathrm{independent,}  \notag \\
&&\left( \ _{\zeta }\psi _{n},\ ^{\zeta }\psi _{n}\right) =0,\ \ n\in \Omega
_{3}\ ,\ _{\zeta }\psi _{n}\ \mathrm{and}\ \ ^{\zeta }\psi _{n}\ \mathrm{%
independent.}  \label{i9}
\end{eqnarray}%
We denote\emph{\ }the corresponding quantum numbers by $n_{k}$,\ so that $%
n_{k}\in \Omega _{k}$. Then we identify:
\begin{eqnarray}
&&\mathrm{in-solutions:\ }_{+}\psi _{n_{1}},\ ^{-}\psi _{n_{1}};\mathrm{\ }%
_{-}\psi _{n_{5}},\ ^{+}\psi _{n_{5}};\ \ _{-}\psi _{n_{3}},\ ^{-}\psi
_{n_{3}}\ ,  \notag \\
&&\mathrm{out-solutions:\ }_{-}\psi _{n_{1}},\ ^{+}\psi _{n_{1}};\ _{+}\psi
_{n_{5}},\ ^{-}\psi _{n_{5}};\ \ _{+}\psi _{n_{3}},\ ^{+}\psi _{n_{3}},
\label{in-out}
\end{eqnarray}

We decompose the Heisenberg operator $\hat{\Psi}\left( X\right) $ in two
sets of solutions $\left\{ \ _{\zeta }\psi _{n}\left( X\right) \right\} $
and $\left\{ \ ^{\zeta }\psi _{n}\left( X\right) \right\} $ of the Dirac
equation (\ref{e1}) complete on the $t=\mathrm{const}$ hyperplane.
Operator-valued coefficients in such decompositions do not depend on
coordinates. Our division of the quantum numbers $n$ in five ranges, implies
the representation for $\hat{\Psi}\left( X\right) $ as a sum of five
operators $\hat{\Psi}_{i}\left( X\right) $, $i=1,2,3,4,5$,%
\begin{equation}
\hat{\Psi}\left( X\right) =\sum_{i=1}^{5}\hat{\Psi}_{i}\left( X\right) .
\label{2.20}
\end{equation}

For each of three operators $\hat{\Psi}_{i}\left( X\right) $,$\ i=1,3,5,$
there exist two possible decompositions according to the existence of two
different complete sets of solutions with the same quantum numbers $n$ in
the ranges $\Omega _{1}$, $\Omega _{3}$, and $\Omega _{5}$,%
\begin{eqnarray}
&&\hat{\Psi}_{1}\left( X\right) =\sum_{n_{1}}\mathcal{M}_{n_{1}}^{-1/2}\left[
\ _{+}a_{n_{1}}(\mathrm{in})\ _{+}\psi _{n_{1}}\left( X\right) +\
^{-}a_{n_{1}}(\mathrm{in})\ ^{-}\psi _{n_{1}}\left( X\right) \right]  \notag
\\
&&\ =\sum_{n_{1}}\mathcal{M}_{n_{1}}^{-1/2}\left[ \ ^{+}a_{n_{1}}(\mathrm{out%
})\ ^{+}\psi _{n_{1}}\left( X\right) +\ _{-}a_{n_{1}}(\mathrm{out})\
_{-}\psi _{n_{1}}\left( X\right) \right] ,  \notag \\
&&\hat{\Psi}_{3}\left( X\right) =\sum_{n_{3}}\mathcal{M}_{n_{3}}^{-1/2}\left[
\ ^{-}a_{n_{3}}(\mathrm{in})\ ^{-}\psi _{n_{3}}\left( X\right) +\
_{-}b_{n_{3}}^{\dagger }(\mathrm{in})\ _{-}\psi _{n_{3}}\left( X\right) %
\right]  \notag \\
&&\ =\sum_{n_{3}}\mathcal{M}_{n_{3}}^{-1/2}\left[ \ ^{+}a_{n_{3}}(\mathrm{out%
})\ ^{+}\psi _{n_{3}}\left( X\right) +\ _{+}b_{n_{3}}^{\dagger }(\mathrm{out}%
)\ _{+}\psi _{n_{3}}\left( X\right) \right] ,  \notag \\
&&\hat{\Psi}_{5}\left( X\right) =\sum_{n_{5}}\mathcal{M}_{n_{5}}^{-1/2}\left[
\ ^{+}b_{n_{5}}^{\dag }(\mathrm{in})\ ^{+}\psi _{n_{5}}\left( X\right) +\
_{-}b_{n_{5}}^{\dag }(\mathrm{in})\ _{-}\psi _{n_{5}}\left( X\right) \right]
\notag \\
&&\ =\sum_{n_{5}}\mathcal{M}_{n_{5}}^{-1/2}\left[ \ _{+}b_{n_{5}}^{\dag }(%
\mathrm{out})\ _{+}\psi _{n_{5}}\left( X\right) +\ ^{-}b_{n_{5}}^{\dag }(%
\mathrm{out})\ ^{-}\psi _{n_{5}}\left( X\right) \right] .  \label{2.23}
\end{eqnarray}

There may\ exist only one complete set of solutions with the same quantum
numbers $n_{2}$ and $n_{4}$. Therefore, we have only one possible
decomposition for each of the\ two operators $\hat{\Psi}_{i}\left( X\right)
, $ $i=2,4$,%
\begin{equation}
\hat{\Psi}_{2}\left( X\right) =\sum_{n_{2}}\mathcal{M}%
_{n_{2}}^{-1/2}a_{n_{2}}\psi _{n_{2}}\left( X\right) ,\ \ \hat{\Psi}%
_{4}\left( X\right) =\sum_{n_{4}}\mathcal{M}_{n_{4}}^{-1/2}b_{n_{4}}^{%
\dagger }\psi _{n_{4}}\left( X\right) .  \label{2.21}
\end{equation}

We interpret all $a$ and $b\ $as annihilation and all $a^{\dag }$ and $%
b^{\dag }$ as creation operators. All $a$ and $a^{\dag }$ are interpreted\
as describing electrons and all $b$ and $b^{\dag }$ as describing positrons.
All the operators labeled by the argument \textrm{in} are interpreted\ as
\textrm{in}-operators, whereas all the operators labeled by the argument
\textrm{out} as \textrm{out}-operators. This identification is confirmed by
a detailed mathematical and physical analysis of solutions of the Dirac
equation with subsequent QFT analysis of correctness of such an
identification in Ref. \cite{GavGit15}.

Taking into account the orthogonality and orthonormalization relations, we
find that the standard anticommutation relations for the Heisenberg operator
(\ref{2.20}) yield the standard anticommutation rules for the introduced
creation and annihilation \textrm{in}- or \textrm{out-}operators. Note that
commutation relations between sets of \textrm{in} and \textrm{out}-operators
follow from the linear canonical transformation that relates \textrm{in} and
\textrm{out}-operators.

We define two vacuum vectors $\left\vert 0,\mathrm{in}\right\rangle $ and $%
\left\vert 0,\mathrm{out}\right\rangle $, one of which is the\ zero-vector
for all \textrm{in}-annihilation operators and the other is zero-vector for
all $\mathrm{out}$-annihilation operators. Besides, both vacua are
zero-vectors for the annihilation operators $a_{n_{2}}$ and $b_{n_{4}}$.
Thus, we have%
\begin{eqnarray}
&&\ _{+}a_{n_{1}}(\mathrm{in})\left\vert 0,\mathrm{in}\right\rangle =\
^{-}a_{n_{1}}(\mathrm{in})\left\vert 0,\mathrm{in}\right\rangle =0,  \notag
\\
&&\ _{-}b_{n_{5}}(\mathrm{in})\left\vert 0,\mathrm{in}\right\rangle =\
^{+}b_{n_{5}}(\mathrm{in})\left\vert 0,\mathrm{in}\right\rangle =0,  \notag
\\
&&\ ^{-}a_{n_{3}}(\mathrm{in})\left\vert 0,\mathrm{in}\right\rangle =\
_{-}b_{n_{3}}(\mathrm{in})\left\vert 0,\mathrm{in}\right\rangle =0,  \notag
\\
&&\ \ a_{n_{2}}\left\vert 0,\mathrm{in}\right\rangle =b_{n_{4}}\left\vert 0,%
\mathrm{in}\right\rangle =0,  \label{cq8a}
\end{eqnarray}

and%
\begin{eqnarray}
&&\ _{-}a_{n_{1}}(\mathrm{out})\left\vert 0,\mathrm{out}\right\rangle =\
^{+}a_{n_{1}}(\mathrm{out})\left\vert 0,\mathrm{out}\right\rangle =0,  \notag
\\
&&\ _{+}b_{n_{5}}(\mathrm{out})\left\vert 0,\mathrm{out}\right\rangle =\
^{-}b_{n_{5}}(\mathrm{out})\left\vert 0,\mathrm{out}\right\rangle =0,  \notag
\\
&&\ _{+}b_{n_{3}}(\mathrm{out})\left\vert 0,\mathrm{out}\right\rangle =\
^{+}a_{n_{3}}(\mathrm{out}\left\vert 0,\mathrm{out}\right\rangle =0,  \notag
\\
&&\ \ a_{n_{2}}\left\vert 0,\mathrm{out}\right\rangle =b_{n_{4}}\left\vert 0,%
\mathrm{out}\right\rangle =0\ .  \label{cq8b}
\end{eqnarray}

One can verify that the introduced vacua have minimum (zero by definition)
kinetic energy and zero electric charge and all the excitations above the
vacuum have positive energies. Then we postulate that the state space of the
system under consideration is the Fock space constructed, say, with the help
of the vacuum $\left\vert 0,\mathrm{in}\right\rangle $ and the corresponding
creation operators. This Fock space is unitarily equivalent to the\ other
Fock space constructed with the help of the vacuum $\left\vert 0,\mathrm{out}%
\right\rangle $ and the corresponding creation operators if the total number
of particles created by the external field is finite.

Because any annihilation operators with quantum numbers $n_{i}$
corresponding to different $i$ anticommute between themselves, we can
represent the introduced vacua as tensor products of the corresponding vacua
in the five ranges,%
\begin{equation}
\left\vert 0,\mathrm{in}\right\rangle =\sideset{}{^{\,\lower1mm\hbox{$%
\otimes$}}}\prod\limits_{i=1}^{5}\left\vert 0,\mathrm{in}\right\rangle
^{\left( i\right) }\ ,\ \ \left\vert 0,\mathrm{out}\right\rangle =%
\sideset{}{^{\,\lower1mm\hbox{$\otimes$}}}\prod\limits_{i=1}^{5}\left\vert 0,%
\mathrm{out}\right\rangle ^{\left( i\right) }\ ,  \label{2.29}
\end{equation}%
where the partial vacua $\left\vert 0,\mathrm{in}\right\rangle ^{\left(
i\right) }$ and $\left\vert 0,\mathrm{out}\right\rangle ^{\left( i\right) }$
obey relations (\ref{cq8a}) and (\ref{cq8b}) for any $n_{i}$ and $\zeta .$

It follows from relations (\ref{cq8a}) and (\ref{cq8b}) that the partial
vacua are stable in $\Omega _{i}$, $i=1,2,4,5$,
\begin{equation}
\left\vert 0,\mathrm{in}\right\rangle ^{\left( i\right) }=\left\vert 0,%
\mathrm{out}\right\rangle ^{\left( i\right) },\ \ i=1,2,4,5.  \label{2.31}
\end{equation}%
Then the total vacuum-to-vacuum transition amplitude $c_{v}$ is due to the
partial vacuum-to-vacuum transition amplitude formed in $\Omega _{3}$,%
\begin{equation}
c_{v}=\langle 0,\mathrm{out}|0,\mathrm{in}\rangle =\ ^{\left( 3\right)
}\langle 0,\mathrm{out}|0,\mathrm{in}\rangle ^{(3)}\ .  \label{cq23}
\end{equation}

The differential mean numbers of electrons and positrons from
electron-positron pairs created are equal:%
\begin{eqnarray}
&&N_{n}^{a}\left( \mathrm{out}\right) =\left\langle 0,\mathrm{in}\left\vert
\ ^{+}a_{n}^{\dagger }(\mathrm{out})\ ^{+}a_{n}(\mathrm{out})\right\vert 0,%
\mathrm{in}\right\rangle =\left\vert g\left( _{-}\left\vert ^{+}\right.
\right) \right\vert ^{-2},  \notag \\
&&N_{n}^{b}\left( \mathrm{out}\right) =\left\langle 0,\mathrm{in}\left\vert
\ _{+}b_{n}^{\dagger }(\mathrm{out})\ _{+}b_{n}(\mathrm{out})\right\vert 0,%
\mathrm{in}\right\rangle =\left\vert g\left( _{+}\left\vert ^{-}\right.
\right) \right\vert ^{-2},  \notag \\
&&N_{n}^{\mathrm{cr}}=N_{n}^{b}\left( \mathrm{out}\right) =N_{n}^{a}\left(
\mathrm{out}\right) ,\ \ n\in \Omega _{3},  \label{7.5}
\end{eqnarray}%
and they present the number of pairs created, $N_{n}^{\mathrm{cr}}$. The
total number of pairs\ created from the vacuum is the sum over the range $%
\Omega _{3}$ of the differential mean numbers $N_{n}^{\mathrm{cr}}$,%
\begin{equation}
N=\sum_{n\in \Omega _{3}}N_{n}^{\mathrm{cr}}=\sum_{n\in \Omega
_{3}}\left\vert g\left( _{-}\left\vert ^{+}\right. \right) \right\vert ^{-2}.
\label{TN}
\end{equation}

Considering the range $\Omega _{3}$, we see that the probabilities of a
particle reflection, a pair creation, and the probability for a vacuum to
remain a vacuum can be expressed via differential mean numbers of created
pairs $N_{n}^{\mathrm{cr}}$,
\begin{eqnarray}
&&P(+|+)_{n^{\prime },n}=|\langle 0,\mathrm{out}|\ ^{+}a_{n^{\prime }}(%
\mathrm{out})\ ^{-}a_{n}^{\dagger }(\mathrm{in})|0,\mathrm{in}\rangle
|^{2}=\delta _{n,n^{\prime }}\frac{1}{1-N_{n}^{\mathrm{cr}}}P_{v}\;,  \notag
\\
&&P(+-|0)_{n^{\prime },n}=|\langle 0,\mathrm{out}|\ ^{+}a_{n^{\prime }}(%
\mathrm{out})\ _{+}b_{n}(\mathrm{out})|0,\mathrm{in}\rangle |^{2}=\delta
_{n,n^{\prime }}\frac{N_{n}^{\mathrm{cr}}}{1-N_{n}^{\mathrm{cr}}}P_{v}\;,
\notag \\
&&P_{v}=|c_{v}|^{2}=\prod\limits_{n}p_{v}^{n},\;\;p_{v}^{n}=\left( 1-N_{n}^{%
\mathrm{cr}}\right) .  \label{cq31}
\end{eqnarray}%
The probabilities for a positron scattering $P(-|-)_{n,n^{\prime }}$ and a
pair annihilation $P\left( 0|-+\right) _{n,n^{\prime }}$ coincide with the
expressions $P(+|+)$ and $P(+-|0)$, respectively.

\section{Asymptotic expansions of WPCFs\label{Ap2}}

Here we list some properties of the WPCFs which are used in the present work.%
\footnote{%
Note that a more detailed description of the properties of the WPCFs can be
found, e.g., in Ref. \cite{BatE53}.}

Asymptotic expansions of WPCFs that correspond to large absolute values of
the argument $\left\vert \xi \right\vert $ have the following form:%
\begin{equation}
D_{\nu }[(1\pm \mathrm{i})\xi ]=e^{\mp \mathrm{i}\xi ^{2}/2}\left( \sqrt{2}%
e^{\pm \mathrm{i}\pi /4}\xi \right) ^{\nu }\left[ 1\mp \mathrm{i}\frac{\nu
\left( 1-\nu \right) }{4\xi ^{2}}+\ldots \right] \quad \mathrm{if}\quad \xi
\geq K,  \label{a1}
\end{equation}%
where $K\gg \max \left\{ 1,\left\vert \nu \right\vert \right\} $. If $\xi <0,
$ we have that%
\begin{eqnarray}
&&D_{\nu }[(1-\mathrm{i})\xi ]=e^{\mathrm{i}\pi \nu }D_{\nu }[(1-\mathrm{i}%
)\left\vert \xi \right\vert ]+\mathrm{i}\frac{\sqrt{2\pi }}{\Gamma (-\nu )}%
e^{\mathrm{i}\pi \nu /2}D_{-\nu -1}[(1+\mathrm{i})\left\vert \xi \right\vert
],  \notag \\
&&D_{-\nu -1}[(1+\mathrm{i})\xi ]=e^{\mathrm{i}\pi \left( \nu +1\right)
}D_{-\nu -1}[(1+\mathrm{i})\left\vert \xi \right\vert ]-\mathrm{i}\frac{%
\sqrt{2\pi }}{\Gamma (\nu +1)}e^{\mathrm{i}\pi \left( \nu +1\right)
/2}D_{\nu }[(1-\mathrm{i})\left\vert \xi \right\vert ],  \label{a2}
\end{eqnarray}%
where $\Gamma (z)$ is the Euler gamma function.

Let $\chi =1$ and $\nu =-1-\rho =\mathrm{i}\lambda /2$. Then, using Eqs.~(%
\ref{a1}) and (\ref{a2}), we obtain the following expansions of the
coefficients $f_{k}^{(+)}(\xi _{l})$, given by Eqs. (\ref{L9}),
\begin{eqnarray}
f_{1}^{(+)}(\xi ) &\approx &e^{-\mathrm{i}\xi ^{2}/2}\left( \sqrt{2}e^{%
\mathrm{i}\pi /4}\xi \right) ^{\nu }2\left[ 1-i\frac{\nu \left( 1-\nu
\right) }{4\xi ^{2}}+O\left( \xi ^{-4}\right) \right] ,  \notag \\
f_{2}^{(+)}(\xi ) &\approx &e^{\mathrm{i}\xi ^{2}/2}\left( \sqrt{2}e^{-%
\mathrm{i}\pi /4}\xi \right) ^{-\nu -1}\left[ -\frac{i}{\xi ^{2}}+O\left(
\xi ^{-4}\right) \right] \quad \mathrm{if}\quad \xi \geq K;  \notag \\
f_{1}^{(+)}(\xi ) &\approx &-ie^{\mathrm{i}\xi ^{2}/2}\left( \sqrt{2}e^{-%
\mathrm{i}\pi /4}\left\vert \xi \right\vert \right) ^{-\nu -1}e^{-i\pi \nu
/2}\frac{\sqrt{2\pi }}{\Gamma \left( -\nu \right) }\left[ 2+O\left( \xi
^{-2}\right) \right] ,  \notag \\
f_{2}^{(+)}(\xi ) &\approx &-e^{\mathrm{i}\xi ^{2}/2}\left( \sqrt{2}e^{-%
\mathrm{i}\pi /4}\left\vert \xi \right\vert \right) ^{-\nu -1}e^{-i\pi \nu }%
\left[ 2-i\left( \frac{3}{2}\nu +\nu ^{2}\right) \xi ^{-2}+O\left( \xi
^{-4}\right) \right]   \notag \\
&&+e^{-\mathrm{i}\xi ^{2}/2}\left( \sqrt{2}e^{\mathrm{i}\pi /4}\left\vert
\xi \right\vert \right) ^{\nu }e^{-i\pi \nu /2}\frac{\sqrt{2\pi }}{2\Gamma
\left( \nu \right) \xi ^{4}}\quad \mathrm{if}\quad \xi <0,\;\left\vert \xi
\right\vert \geq K.  \label{a3}
\end{eqnarray}

\end{document}